\documentclass[osajnl,twocolumn,showpacs,superscriptaddress,10pt]{revtex4-1} %% use 11pt for Applied Optics
\usepackage{amsmath,amssymb,graphicx}
\usepackage{enumerate}
\usepackage{xcolor}

\begin{document}

\title{The physics of the photothermal detection of single absorbing nano-objects: A review}

\author{Markus Selmke}\email{Corresponding author: markus.selmke@gmx.de}
%\author{Andr\'e Heber}
\author{Frank Cichos}
\affiliation{Universit\"at Leipzig, Experimental physics I, molecular nanophotonics, Linn\'estr.\ 5, 04103 Leipzig, Germany}

\begin{abstract}
The literature on the theory of the photothermal imaging of single nano objects is reviewed. Several models have been devised to describe the signals magnitude, phase and shape in dependence on the experimental and sample parameters (particle position and optical properties, laser beams and their offset, modulation frequency, (thermo-)refractive coefficients, thermal constants). The benefits and limitations of these models are summarized allowing the proper choice of a framework for future investigations using photothermal microscopy of single absorbers.
\end{abstract}

\maketitle %% required

\section{Introduction}
Photothermal microscopy has found its way into the standard repertoire of methods aimed at the investigation of the optical properties and specifically the absorption cross-section of single nano objects. The thermal wave further allows the assessment of thermal transport and phase-transitions on the nanoscale in the host-medium and potentially the particle itself. A growing number of studies use this highly sensitive and selective method already, see for instance the recent review article by L.\ Cognet \cite{Brahim2012}. Recent experimental progress includes the detection of single molecules, quantitative ab-initio spectroscopy \cite{SelmkeACSNano} and superresolution photothermal microscopy utilizing non-linearity \cite{Nedosekin} or pupil filters \cite{Miyazaki2014}.

The first attempt of a description of the signal was already made in 2006 by the same group which invented its most sensitive variant, the so-called photothermal heterodyne imaging. In this scheme, a modulated pump beam is focused onto the nano object whereby a thermal lens $n\left(r,t\right)$, eq.\ \ref{eq:nr}, is created in the embedding material.
\begin{equation}
n\left(r,t\right)=n_0 + \Delta n\frac{R}{r}\left[1+\exp\left(-\frac{r}{R_{\rm th}}\right)\cos\left(\Omega t- \frac{r}{R_{\rm th}}\right)\right]\label{eq:nr}.
\end{equation}
The contrast $\Delta n=(\partial_{T} n) P_{\rm abs}/4\pi\kappa$ of the thermal lens is determined by the absorbed power $P_{\rm abs}$, the embedding mediums' heat conductivity $\kappa$ and the pump beam's modulation frequency $\Omega$. Together with the mediums' heat-capacity $C$ per unit volume, the latter parameters also determine the exponential decay length $R_{\rm th}=\sqrt{2\kappa/\Omega C}$ of the thermal diffusion wave field. The absorbed power is determined by the  spatial intensity of the pump laser at the particle position $\mathbf{r_p}$ and the particle's absorption cross-section $P_{\rm abs}\approx I_{h}\left(\mathbf{r_p}\right)\sigma_{\rm abs}$. A second laser which is co-aligned with the pump-beam, possibly offset axially by $-\Delta z_f$, probes the thermal lens. Either the transmittance or a reflected part of the probe beam is then imaged onto a photodiode where the relative signal $\Delta P_d\left(t\right)/P_{d}$ is recorded. A lock-in detection allows for an effective background reduction and the high selectivity towards absorbers in an otherwise scattering environment. If the modulation frequency is moderate such that the lens is unscreened over the extent of the probe laser, i.e.\ $R_{\rm th}\left(\Omega\right)\gg \omega_0$, the ideal thermal lens may be assumed for a theoretical evaluation of the signal,
\begin{equation}
n\left(r,t\right)\approx n_0 + \Delta n\frac{R}{r}\left[1+\cos\left(\Omega t\right)\right] \label{eq:nrIdeal},
\end{equation}
i.e.\ a modulated long-ranged $1/r$ profile.

In this article we summarize the progress which has been made in the fundamental understanding of the signal. \textit{Firstly}, we show that only those models which explicitly consider a focused beam are able to give an accurate account of the interference phenomenon. An energy-redistribution of the probe-beam is essential to the PT signal, and as such must be considered. Accordingly, a dispersive signal as a function of the axial placement of the particle in the focus is found, and thereby a vanishing signal occurs if the probe beam is focused directly onto the thermal lens, whereas a maximal signal is attained for some probe-beam offset of about one Rayleigh-range. \textit{Secondly}, we conclude that a purely interferometric interpretation is only applicable for PT microscopy in the high-frequency limit $R_{\rm th}\ll \lambda$. Otherwise, it misses the long-range character of the thermal lens signal in situations where the signal is maximal. Then, both scattering and interference contribute on equal footing to the observed linear dependence on the heating power. Their concurrent role testify the intrinsic character of a lensing action which lies at the heart of PT microscopy. \textit{Thirdly}, the maximum attainable signal will be shown to scale with the refractive index contrast that is induced, i.e.\ with a quantity $\nu\propto \Delta n$.
All three conclusions are in accordance with macroscopic thermal lensing spectroscopy methods and appear quite natural. However, the current literature is at odds with several of these observations. 
The signal in the backwards direction appears to follow the forward transmitted signal apart from an overall amplitude factor determined by the retroretlecting interface. This suggests the same lensing mechanism at work such that this article will concentrate on the transmission detection scheme.

\begin{widetext}

\begin{figure}[t]
\centerline{\includegraphics [width=\columnwidth]{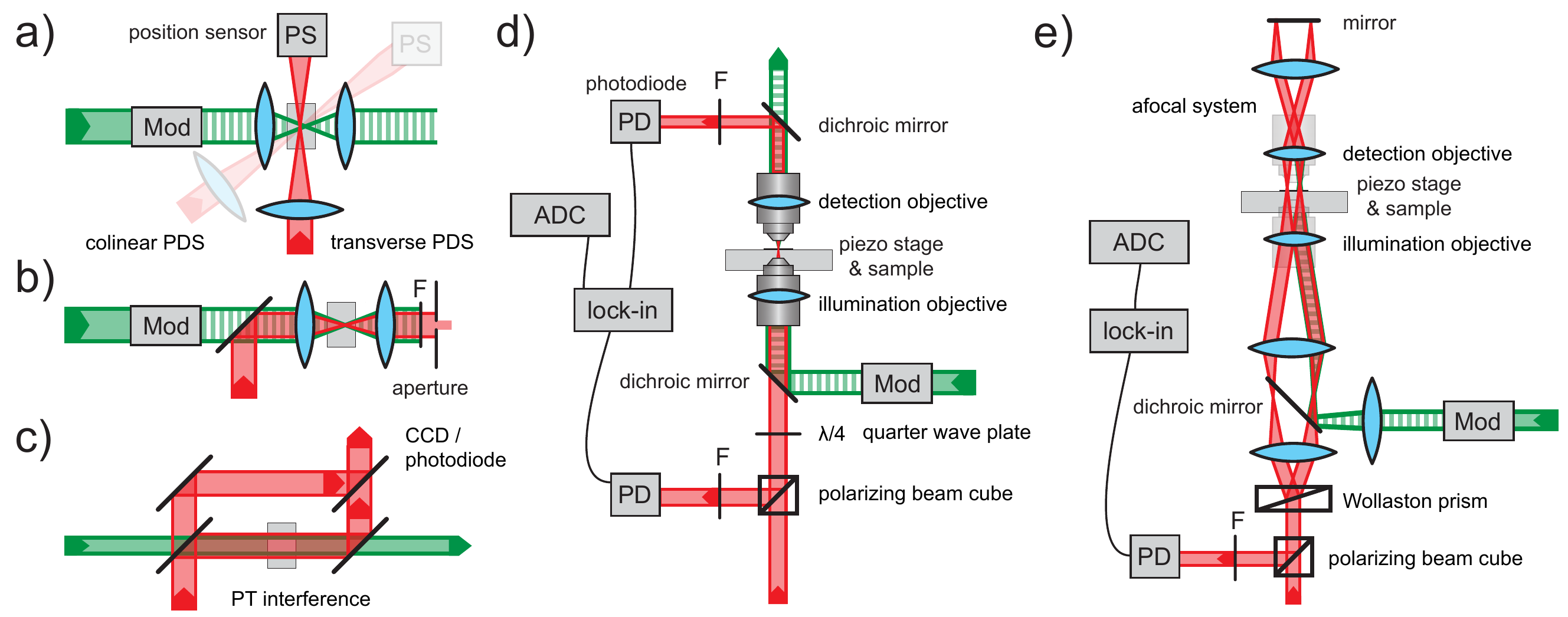}}
\caption{Macroscopic thermal spectroscopy variants methods \cite{Bialkowski}: \textbf{a)} Photothermal deflection (mirage) spectroscopy (coaxial and lateral) which probes the gradient $\nabla n$ \cite{Jackson1981}. \textbf{b)} Thermal lens (TL) spectroscopy which probes the curvature $\partial^2 n/\partial r_{\perp}^2$ \cite{Jackson1981}. \textbf{c)} Photothermal interference spectroscopy which is sensitive to a phase advance $\Delta \chi=\int \! k\, n\mathrm{d}s$. These techniques use thin macroscopic samples and weakly focused beams and the signal is proportional to the induced refractive index contrast $\Delta n$. Experimental setup schemes for thermal lens microscopy, PT DIC and PT lens microscopy \cite{Cognet2008Review}: \textbf{d)} PT differential interference contrast (PIC) microscopy due to M.\ Orrit and co-workers \cite{Boyer2002}. \textbf{e)} PT heterodyne imaging introduced by B.\ Lounis and co-workers \cite{BerciaudCognet2004,Berciaud2006}. The forward channel is an extension of the thermal lens microscope by M.\ Harada and T.\ Kitamori \cite{Harada1993AnalChem,Kitamori2000}.\label{SetupSchemesPT}}\label{Fig:Schemes}
\end{figure}

\end{widetext}

The following models have been proposed: 
\begin{enumerate}[A)]
\item Heterodyne imaging theory \cite{Berciaud2006,Miyazaki2014Optimal}
\item Equivalent dipole model \cite{Gaiduk2010, Paulo2009}
\item Generalized Lorenz-Mie theory \cite{NanoLensDiff,SelmkeACSNano} %(multilayered scatterer)
\item Photonic Rutherford scattering \cite{SelmkePRL2013, SelmkeAmJPhys2013}
\item Fresnel Diffraction \cite{NanoLensDiff} / PIC \cite{Cognet2003}
\item Gaussian ABCD method \cite{SelmkeABCD}
\end{enumerate}
Their capabilities, benefits and limitations regarding the evaluation and characteristics of the relative PT signal, eq.\ \eqref{PTsignal}, shall be given.
\begin{equation}
\Phi=\frac{\Delta P_d}{P_{\rm inc}}\label{PTsignal},
\end{equation}
wherein $\Delta P_d$ is the change in transmission upon heating via the pump laser beam. The signal will have two components, one of which is in phase ($\Phi_{\rm cos}$) with the heating and another which is out-of-phase ($\Phi_{\rm sin}$). The expression Eq.\ \eqref{PTsignal} thereby decomposes into two terms. The absolute value of the signal reads $\Phi=\sqrt{\Phi_{\sin}^2+\Phi_{\cos}^2}$. In the forward detection scheme, the transmitted power is collected up to an angle $\theta_{\rm max}=\arcsin\left({\rm NA}_d/n_0\right)$ determined by the collection objective's numerical aperture ${\rm NA}_d$ and the embedding medium (assuming a dry objective). The illumination objective in turn influences the spatial profile of the incident probe and heating fields.
\begin{widetext}

\begin{figure}[bth]
\centerline{\includegraphics [width=\textwidth]{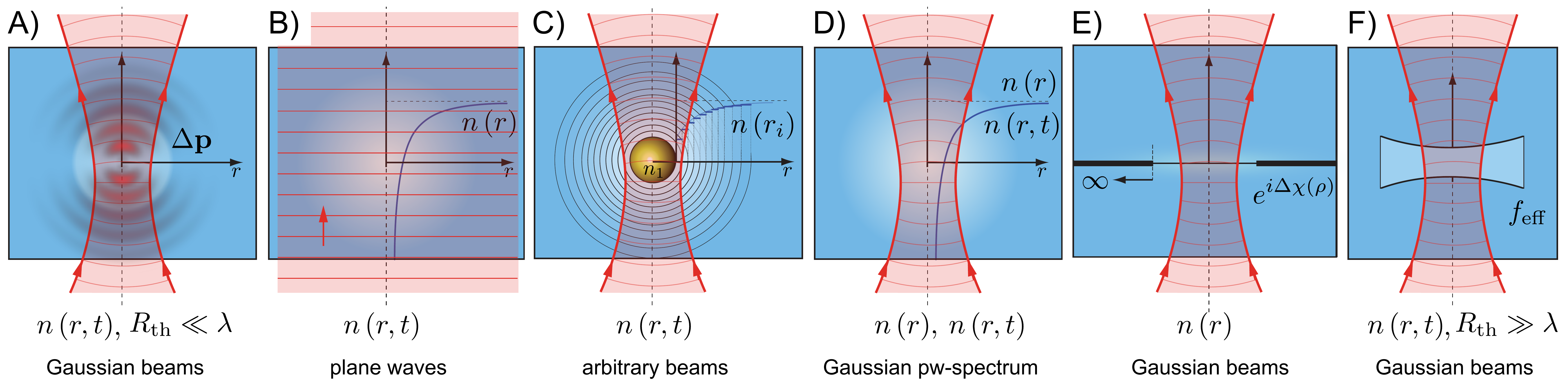}}
\caption{Schematic of the available models for PT single particle microscopy.}\label{Fig:Rutherford}
\end{figure}

\end{widetext}

\section{The equivalent Dipole model (A)}
The tempting conceptual similarity to the scattering by a single point-like dipole was suggested \cite{Gaiduk2010, Paulo2009} to provide an explanation of the PT signal via a purely interferometric mechanism. Along the lines of scattering by a perturbation in the dielectric constant $\epsilon=\epsilon_0 n^2$ which acts as an induced dipole-source \cite{JacksonBook} of strength 
\begin{equation}
\delta \mathbf{p}=\delta \epsilon \delta V\mathbf{E^{i}}=2\epsilon_0 n_0 \delta n \delta V \mathbf{E^{i}}\equiv \epsilon \alpha_n \mathbf{E^{i}},
\end{equation}
the forward signal would then correspond to the interference of the radiated dipole-field with the incidence beam. Such a treatment is expected to be valid for high frequencies such that the scatterer is small, i.e.\ for $R_{\rm th}\left(\Omega\right)\ll \lambda$. This is the Rayleigh limit. It provides a qualitative description of the signal-to-noise ratio for the backwards detection scheme \cite{Gaiduk2010}. Building up on these ideas, a brief derivation of the transmission signal in the narrow forward direction shall be given for an induced dipole \cite{SelmkeApplPhysLett2014}.

%\begin{figure}[b]
%\centerline{\includegraphics [width=\columnwidth]{Fig_Dipole.pdf}}
%\caption{Schematic of the interference of a paraxial Gaussian beam with an induced dipole $\delta \mathbf{p}$.}\label{Fig:ScanZX}
%\end{figure}

The resulting interference signal for transmission is $\left[|\mathbf{E^s}+\mathbf{E^i}|^2-|\mathbf{E^i}|^2\right]/|\mathbf{E^i}|^2 \rightarrow 2\,\mathfrak{R}(\mathbf{E^{i*}}\mathbf{E^s})/|\mathbf{E^i}|^2$, which may be evaluated for a Gaussian beam. For the strict forward direction the evaluation simplifies considerably. In a beam-waist centered coordinate-system the electric field amplitude of the incidence beam is:
\begin{equation}
\mathbf{E^i}\left(\rho,z\right)\approx-\mathbf{\hat{e}_x} \frac{E_0 \exp\left(-ikz\right)}{1-i z/z_R}\exp\left(\frac{-\rho^2/\omega_0^2}{1-iz/z_R}\right),
\end{equation}
with the wave-vector $k=n_0 2\pi/\lambda$, Rayleigh-range $z_R=k\omega_0^2/2$ and beam-waist $\omega_0$. The Gouy phase $\phi_G=\arctan\left(z_p/z_R\right)$ is contained in the exponential prefactor. In the far field and on the optical axis this amounts to the following field, now written in a particle centered coordinate-system ($z_p>0$ corresponds to a particle behind the beam-waist in propagation direction):
\begin{equation}
\mathbf{E^i}\left(z\right)\rightarrow -\mathbf{\hat{e}_x} E_0 z_R \frac{\exp\left(-ik\left[z+z_p\right]+i\pi/2\right)}{z}.
\end{equation}
The induced dipole moment depends on the gaussian beam's field amplitude $\mathbf{E^i}\left(0,z_p\right)$ at the position of the particle, i.e.\
\begin{equation}
\delta \mathbf{p}= -\mathbf{\hat{e}_x} \epsilon \alpha_n E_0 \exp\left(-ikz_p\right) \frac{1}{1-i z_p/z_R}
\end{equation}
such that in the near-forward direction the interference with the radiated dipole far field ($kr\gg 1$) $\mathbf{E^s}\rightarrow \left[k^2/4\pi\epsilon\right] \left(\mathbf{\hat{e}_{\mathbf{r}}}\times \delta\mathbf{p}\right)\times \mathbf{\hat{e}_{\mathbf{r}}} \exp\left(-ikr\right)/r$ can be evaluated. It should be noted, that it is at this point that one could come to the conclusion that the signal should be proportional to $n_0 \delta n$, as concluded in the first theoretical work of S. Berciaud et al.\ \cite{Berciaud2006} and other works \cite{Gaiduk2010}. Incorporating the focused beam and the effective polarizability $\alpha_n=2n_0^{-1}\delta n \delta V$, however, we find for the interference signal \cite{SelmkeApplPhysLett2014}
\begin{equation}
\Phi  = \frac{k}{\pi\omega_0^2}\mathfrak{R}\left(\frac{-i \alpha_n}{1-i z_p/z_R}\right)= \frac{4\delta n \delta V}{\lambda\omega_0^2}\frac{z_p/z_R}{1+z_p^2/z_R^2}.\label{Eq:PTHF}
\end{equation}
%Considering instead the complex-valued polarisability $\alpha$ of the central absorbing particle in the above expression, the unmodulated background signal due to the particle scattering and interference may be quantified. The resulting axial shape is then a weighted sum of a dispersive and a dip-like shape, belonging to the real and imaginary parts of the polarisability, respectively. The same result may be obtained from the rigorous GLMT in the small-particle limit \cite{SelmkeTransmission2014}. Returning to the time-varying photothermal contribution we find the following: 
which is valid for the assumed single effective dipole of real-valued polarizability. For some fixed offset $z_p\ne 0$ the signal is non-zero and proportional to $\Phi \propto \delta n /z_R \propto \Delta n/n_0$. This is the optical contrast that must be the signal-determining quantity for a lensing / interference situation. It should also be noted that this behaviour is typical for thermal lensing, see the book of Bialkowski \cite{Bialkowski}. The maximum signal attainable, in any model for that matter, occurs at a finite offset $z_p\approx z_R$ and is proportional to the perturbation ${\rm max}\left(\Phi\right)\propto \Delta n\propto \left[\partial_T n\right]/\kappa$ itself. This could be called the photothermal figure of merit characterizing the embedding medium, cf. \cite{Gaiduk2010}. These proportionalities remain for arbitrary collection angles and underlying dipole distributions (thermal lens profiles).

%However, it is rather futile to attempt to verify either proportionality, $\times n_0$ or $\times n_0^{-1}$, since aberrations sensitively depend on the embedding medium and dominate such effect. Changing $n_0$ in a controlled manner independent of $\omega_0$ and $z_p$ is difficult if not impossible.

Since the above treatment will only be valid for high frequencies at which the modulated refractive index perturbation decays exponentially with $R_{\rm th}\ll \lambda$, the quantity $\delta n \delta V$ in Eq.\ \eqref{Eq:PTHF} may be approximated via $\int \Delta n\left(\mathbf{r},t\right) \mathrm{d}\mathbf{r}=2\pi R \Delta n R_{\rm th}^2 \sin\left(\Omega t\right)$, where the integration is over all space and using the instantaneous thermal lens Eq.\ \eqref{eq:nr} at some fixed time $t$. One finds \cite{SelmkeApplPhysLett2014}:
\begin{equation}
\Phi_{\rm sin}  =  \frac{4 P_{0,h} \sigma_{\rm abs} \left[\partial_T n\right]}{\pi \omega_{0,h}^2 \kappa \lambda}\left[\frac{R_{\rm th}^2}{\omega_0^2}\right]\frac{z_p/z_R}{1+z_p^2/z_R^2}.\label{Eq:PTHFsin}
\end{equation}
As a result, this model predicts a decay of a signal which is out-of-phase $\Phi\approx \Phi_{\rm sin}$ and $\propto \Omega^{-1}$, a result found early for plane-wave interference \cite{Berciaud2006}. The above result gives a good quantitative approximation for the near-forward signal. For finite collection angles the signal decays to zero over an angular range of twice the probing beam's angle of divergence $\theta_{\rm div}$ \cite{SelmkeTransmission2014}. However, extrapolating the recorded signal to $\theta_{\rm max}/\theta_{\rm div}\ll 1$ allows the quantitative comparison to the above Eq.\ \eqref{Eq:PTHFsin}.

The dispersive signal shape may be understood as characteristic feature for an object which redistributes the energy flux of the probe-beam. Even for larger detection apertures, the signal remains dispersive. However, for increasing collection angles the magnitude decreases monotonically. For collection angles larger than twice the beam's angle of divergence, $\theta_{\rm div}=2/k\omega_0$, the signal vanishes, indicating that the energy is only redistributed in the forward direction due to interference and a complete collection then results in a zero signal \cite{SelmkeTransmission2014}. A similar dispersive signature is also found for a lens of focal length $f$, where the signal reads $\phi=2z_p/f$. The details of the dispersive shape depend on the mechanism of the redistribution itself. The ideal thermal lens will behave as $\Phi\propto \Delta n \arctan\left(z/z_R\right)$ instead, also showing a zero-crossing. The discussed limitation of this model to high frequencies prevents a simple extension to the quasi-static scenario of high refractive contrast. Indeed, the signal magnitude is larger for the in-phase contribution at moderate or low frequencies where $R_{\rm th}\sim \lambda$ which the model cannot describe. For larger decay lengths, the full thermal lens must be regarded either as a large (Mie-)scatterer (see model C) or as an extended distribution of dipoles (see model (B)). The transition between both regimes and thereby the transition form $\Phi \propto \left[z/z_R\right]/\left[1+z^2/z_R^2\right]$ to $\Phi \propto \arctan\left(z/z_R\right)$ is smooth. Both axial functions agree up to second order in the relative displacement.

Regarding the effect of the central NP, the nonlocal character of the particle's dependence of its complex polarizability $\alpha$ on the refractive index of the surrounding thermal lens will lead to both in-phase and out-of-phase contributions. Using the general interference expression in Eq.\ \eqref{Eq:PTHF}, it may be checked that the ratio of the changing NP interference due to $|\partial \alpha/\partial n| \Delta n \sim |\alpha| \Delta n$ and the direct PT signal due to the lens is about $\sim R \omega_0^2 /|\alpha| \sim \omega_0^2/R^2\gg 1$ for the case of slow modulation, and about $\sim R R_{\rm th}^2/|\alpha| \sim R_{\rm th}^2/R^2$ for high frequencies (anticipating the result Eq.\ \eqref{eqPhiApproxDiffrOnAxis}). Therefore, only at high frequencies with $R_{\rm th}\left(\Omega\right)\sim R$ will this indirect PT signal due to the modulated particle interference occur.

%\begin{figure}[bth]
%\centerline{\includegraphics [width=\columnwidth]{zScanEvolution.pdf}}
%\caption{Evolution of the shape of z-scans $\Phi\left(z_p\right)$.}\label{Fig:ScanZX}
%\end{figure}

\section{Photothermal Heterodyne Imaging (B)}
The first insightful theory on the details of the signal generation process in PT microscopy was given using the notion of a fluctuating medium following the work of J.B.\ Lastovka. The scattered field is then found from a vector Hertz potential utilizing a polarization density which is proportional to $\left[\partial n/\partial T\right] / n_0$, see loc.\ eq.\ (2) and (6) of Ref.\ \cite{Berciaud2006}. Equivalently, one may consider the vectorial Born approximation, where the effective induced dipole moment density is $\mathrm{d}\mathbf{p}/\mathrm{d} V=\mathrm{d} \epsilon\, \mathbf{E^{i}} = 2\epsilon_0 n_0 \Delta n\left(\mathbf{r}\right) \mathbf{E^{i}}$. The radiated fields of such an extended distribution of dipoles interfere with the incident beam. However, it suffices to consider the scalar scattering scenario in the Born approximation instead to capture the interference characteristics.

The inhomogeneous Helmholtz equation for the field amplitudes,
\begin{equation}
\nabla^2 U\left(\mathbf{r}\right)+k^2 \left[\frac{n\left(\mathbf{r}\right)}{n_0}\right]^2 U\left(\mathbf{r}\right)=0\label{eq:HE}
\end{equation}
may be rewritten as a scalar homogeneous HE with a source term, $\left[\nabla^2+k^2\right]U\left(\mathbf{r}\right)=-4\pi F\left(\mathbf{r}\right) U\left(\mathbf{r}\right)$, introducing the so-called scattering potential $F\left(\mathbf{r}\right)=k^2\left[n^2\left(\mathbf{r}\right)/n_0^2 - 1\right]/4\pi$:
\begin{equation}
F\left(\mathbf{r}\right) \approx \frac{-k \nu}{2\pi}\left[\frac{1}{r} + \frac{\exp\left(-r/R_{\rm th}\right)}{r}\cos\left(\Omega t-\frac{r}{R_{\rm th}}\right)\right] \label{eq:ScattPot}
\end{equation} 
with $\nu=-k R \Delta n/n_0$ and terms of order $\nu^2$ have been neglected. In the BA the total scalar electric field is decomposed into the sum of an incident plane wave with a medium wave-vector of magnitude $k=k_0 n_0$ and in the $z$-direction, and a spherical scattered wave according to $U\left(\mathbf{r}\right)/U_0=\exp\left(i \mathbf{k}\cdot\mathbf{r}\right) + e^{ik r} r^{-1} f\left(\theta\right)$. The amplitude of the scattered spherical wave is related to the scattering potential. For a spherically symmetric scattering potential, the absolute value $q=2k\sin\left(\theta/2\right)$ of the momentum transfer vector determines the scattering amplitude in any direction specified by the polar angle $\theta$:
\begin{equation}
 f^{\rm BA}\left(\theta\right)=\frac{4\pi}{q} \int F\left(r'\right)\sin\left(qr'\right)r' \mathrm{d}r'\label{eq:fBA}
\end{equation} 
The first order BA approximates the total field inside the scattering potential region by the unperturbed incident field. For the thermal lens, and expanding the cosine in eq.\ \eqref{eq:ScattPot}, one finds \cite{SelmkeApplPhysLett2014}
\begin{equation}
f^{\rm BA}_\Omega\left(\theta,t\right)=f^{\rm BA}_C + \cos\left(\Omega t\right)f^{\rm BA}_{\cos} +\sin\left(\Omega t\right)f^{\rm BA}_{\sin},\label{eq:fomega}
\end{equation}
with the individual amplitudes
\begin{align}
f^{\rm BA}_C\left(\theta\right)&=\frac{-2k\nu}{q}\!\int_{0}^{\infty}\!\sin\left(qr\right)\mathrm{d}r\label{eq:fBA}\\
f^{\rm BA}_{\cos}\left(\theta\right)&=\frac{-2R_{\rm th}^4k^3\nu\sin^2\left(\frac{\theta}{2}\right)}{1+4R_{\rm th}^4k^4 \sin^4\left(\frac{\theta}{2}\right)},\\
f^{\rm BA}_{\sin}\left(\theta\right)&=\frac{-R_{\rm th}^2 k \nu}{1+4R_{\rm th}^4k^4 \sin^4\left(\frac{\theta}{2}\right)}.\label{eq:fCosSon}
\end{align}
From the literature on Coulomb scattering, it is known that the scattering amplitude eqn.\ \eqref{eq:fBA} for the unmodulated background scattering potential is indeterminate in the BA. However, using an exponential Yukawa-type screening and considering the limit of an unscreened potential only after the amplitude evaluation, the following value may be assigned to it:
\begin{equation}
f^{\rm BA}_C\left(\theta\right)=-\frac{\nu}{2k}\sin^{-2}\left(\theta/2\right).\label{eq:fC}
\end{equation}
\begin{figure}[bth]
\centerline{\includegraphics [width=\columnwidth]{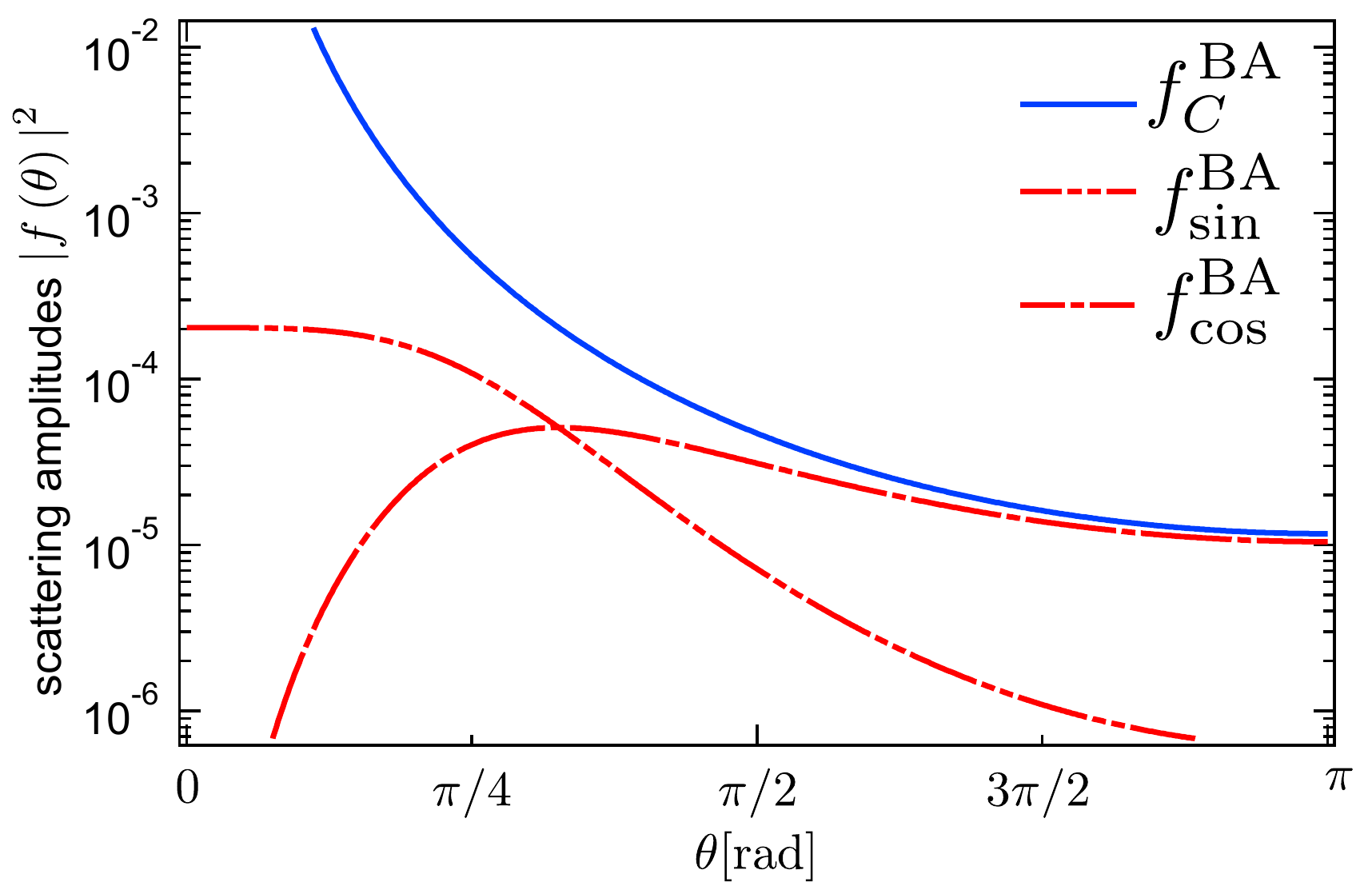}}
\caption{Exemplary plane-wave scattering amptitudes $f\left(\theta\right)$ by the thermal wave $n\left(r,t\right)$ for $k R_{\rm th}=1.44$ and $\nu=10^{-3}$.}\label{Fig:ScanZX}
\end{figure}
Already at this point it becomes clear, that a plane-wave treatment of the thermal lens is problematic although a plane wave well approximates a Gaussian beam near its focus. Intrinsic to this treatment is the neglect of the Gouy phase and the local character of the probing. The interference of an induced dipole with a focused beam as carried out above illustrates their importance vividly. In Ref.\ \cite{Berciaud2006}, however, the evaluation of the interference signal was suggested to be tantamount to the inclusion of a factor $\left[1+\cos\left(\theta\right)\right]$ in evaluating the far-field interference. The same difficulties remain in the time-modulated part of the scattering amplitude even if one accepts this reasoning. For this we consider the analytical integration in the entire forward angular domain, and normalize accordingly by $3|U_0|^2/2$, i.e.\ 
\begin{align}
\Phi_{\cos}=& \displaystyle \frac{2k}{3}\int_{0}^{\pi/2}f^{\rm BA}_{\cos}\left[1+\cos\left(\theta\right)\right]\sin\left(\theta\right)\mathrm{d}\theta,\label{SignalBerciaudF}\\
=& -\frac{2\nu}{3} \left[\frac{\arctan\left(k^2 R_{\rm th}^2\right)}{k^2 R_{\rm th}^2}+\ln\left(1+k^4 R_{\rm th}^4\right) - 1\right]\nonumber\\
\Phi_{\sin} =& \displaystyle \frac{2k}{3}\int_{0}^{\pi/2}f^{\rm BA}_{\sin}\left[1+\cos\left(\theta\right)\right]\sin\left(\theta\right)\mathrm{d}\theta,\\
=& -\frac{2\nu}{3}\left[2\arctan\left(k^2 R_{\rm th}^2\right) - \frac{\ln\left(1+k^4 R_{\rm th}^4\right)}{2k^2 R_{\rm th}^2}\right].\nonumber
\end{align}
Similar expressions are obtained in the backward direction. The above expressions reproduce the numerical results of what is claimed to be the frequency dependence of the PT signal in Ref.\ \cite{Berciaud2006}. Also for finite collection angle the behaviour remains, although shifted along the ordinate. It is seen, that the in-phase forward signal diverges logarithmically as $\Phi_{\cos}\rightarrow -8\nu\ln\left(k R_{\rm th}\right)/3$ and the out-of-phase signal saturates in the quasi-static limit of moderate or low frequencies where the signal becomes large. These are artefacts due to the assumed plane-wave probing of a problematic infinite potential. For high modulation frequencies ($R_{\rm th} \ll \lambda$) the dependence of the in-phase and out-of-phase contributions are $\Phi_{\cos}\propto -\nu k^4R_{\rm th}^4$ and $\Phi_{\sin}\propto -\nu k^2R_{\rm th}^2$, respectively. 

\begin{figure}[b]
\centerline{\includegraphics [width=\columnwidth]{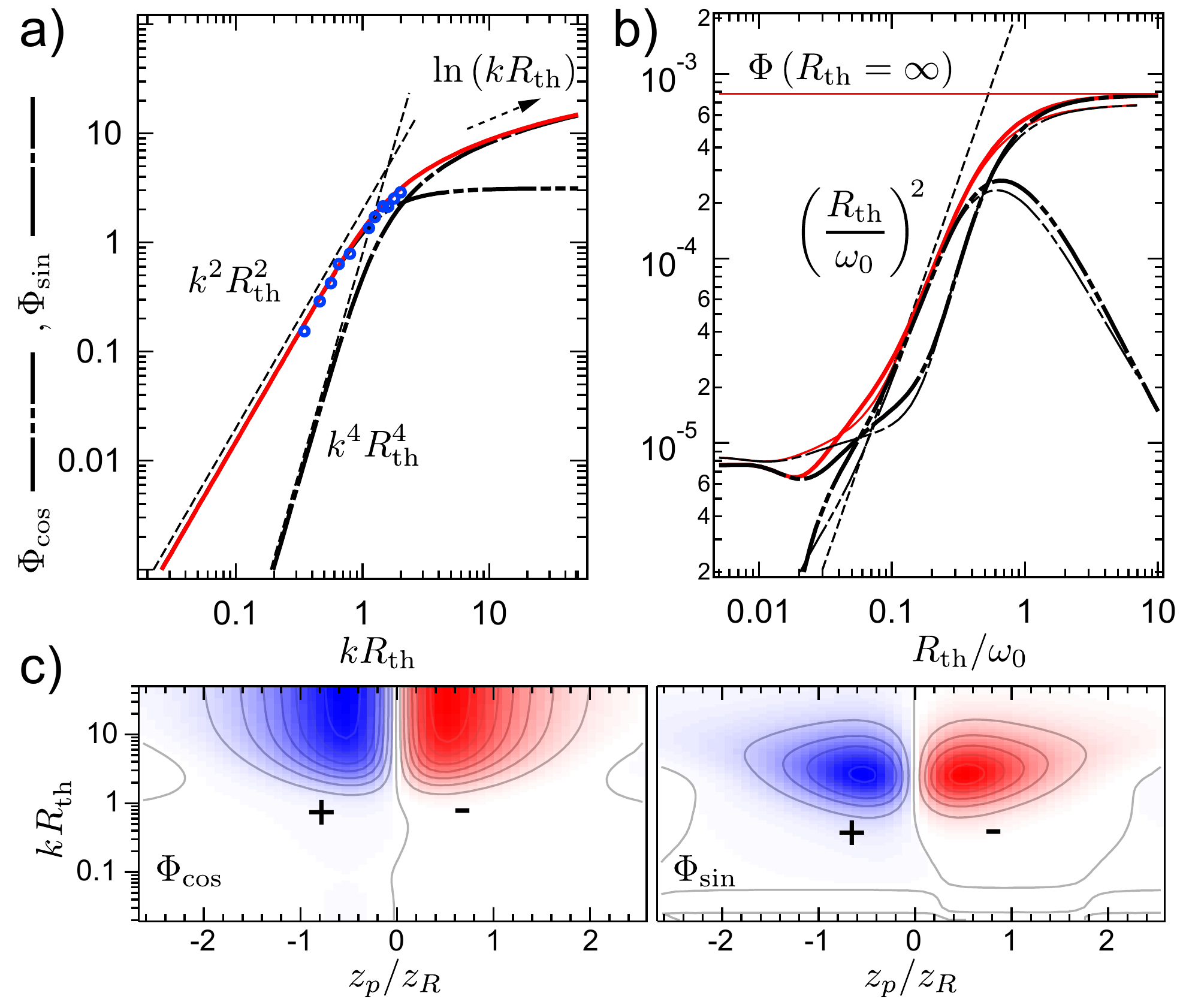}}
\caption{\textbf{a)} Screening length dependence of the rel.\ PT signal, $\Phi\left(R_{\rm th}\right)$ according to the accurate GLMT model ($R=10\,\rm nm$, thick lines: $\omega_0=281\,\rm nm$ and ${\rm NA}_d=0.8$ thin lines $\omega_0=500\,\rm nm$ and ${\rm NA}_d=0.5$) and \textbf{b)} the plane-wave Born approximation theory. \textbf{c)} Axial scans of the in-phase $\Phi_{\cos}\left(z_p\right)$ and out-of-phase $\Phi_{\sin}\left(z_p\right)$ contributions for various scaled thermal diffusion lengths $k R_{\rm th}$.}\label{Fig:Frequency}
\end{figure}

The correct behaviour of the in- and out-of-phase signals, e.g.\ obtained in the GLMT model, systematically differs from these predictions and agrees with experiments, see Fig.\ \ref{Fig:Frequency}. Only the decay of the absolute value $\Phi\propto \nu k^2R_{\rm th}^2\propto 1/\Omega$ of the signal for intermediate frequencies happens to agree for both models. Measurements support the GLMT result and indeed show a peak of the out-of-phase signal for $R_{\rm th}\left(\Omega=\Omega_p\right) \approx \omega_0$. This fact can be used to determine the thermal diffusivity $\kappa/C$ of the embedding host material for a known beam-waist via the peak frequency, i.e.\ $\kappa/C \approx \Omega_p \omega_0^2/2$ \cite{SelmkeApplPhysLett2014}. A very similar protocol has recently been proposed by N.\ J.\ Dovichi et al.\ using a square-pule modulation to image thermal diffusivities via bulk-absoption with a spatial resolution of a few $\mu \rm m$ \cite{Dovichi2010}. The resolution of the thermal diffusivity determination via point-like heat-source is likely also determined by the probe beam waist.

Realizing the inadequacy of the plane-wave formalism for the PT signal for focused beam probing, it appears not very cogent to assume that the backwards signal is correctly described in this model. Unfortunately, so far no coherent quantitative model has been developed for the backwards detected signal.

Returning to the forward signal, an exact treatment of unscreened scattering potential Eq.\ \eqref{eq:ScattPot} is possible and avoids the previously mentioned divergencies. This ansatz leads to the model termed "photonic Rutherford scattering", described in detail in section \ref{sec:photRF} (model D). In combination with the wave-packet formalism this allows a correct description of the signal. Such a correct treatment yields a zero signal for the case assumed here, namely a particle placed in the focus. In contrast, no axial dependence of the signal is contained in Eq.\ \eqref{SignalBerciaudF} and a finite signal is expected. Notably, the plane-wave solution Eq.\ \eqref{eq:fomega} at finite frequencies can be used to describe focused beam scattering \cite{Miyazaki2014Optimal} in a plane-wave expansion of the probing beam, see the comment in that section.

The vectorial character of the incident beam doesn't change these peculiarities. The scattered field of a vectorial electromagnetic plane wave with a polarization vector $\mathbf{\hat{e}_0}$ is connected to the scalar amplitude via $\mathbf{E}^s \left(\mathbf{r}\right) = -E_0\,\mathbf{\hat{e}}_{\mathbf{r}}\times\left[\mathbf{\hat{e}}_{\mathbf{r}}\times\mathbf{\hat{e}_0}\right] e^{ikr} r^{-1} f^{\rm BA}\left(\theta\right)$. Unfortunately, the interference signal cannot be inferred from the above scattering amplitude, say via an effective polarisability $\alpha_n=4\pi f^{\rm BA}\left(0^{\circ}\right)/k^2$. The Gouy phase shift plays a crucial role for focused beam interference and is not included in the evaluation of the scattering amplitude via eq.\ \eqref{eq:fBA}. An extension of the Born approximation to include these effects, avoiding the singularity in the forward direction via the inclusion of a focused beam, is not likely to be simpler than the rigorous GLMT model disused next.

\section{Generalized Lorentz Mie theory (C)}
\begin{widetext}

\begin{figure}[bt]
\centerline{\includegraphics [width=\columnwidth]{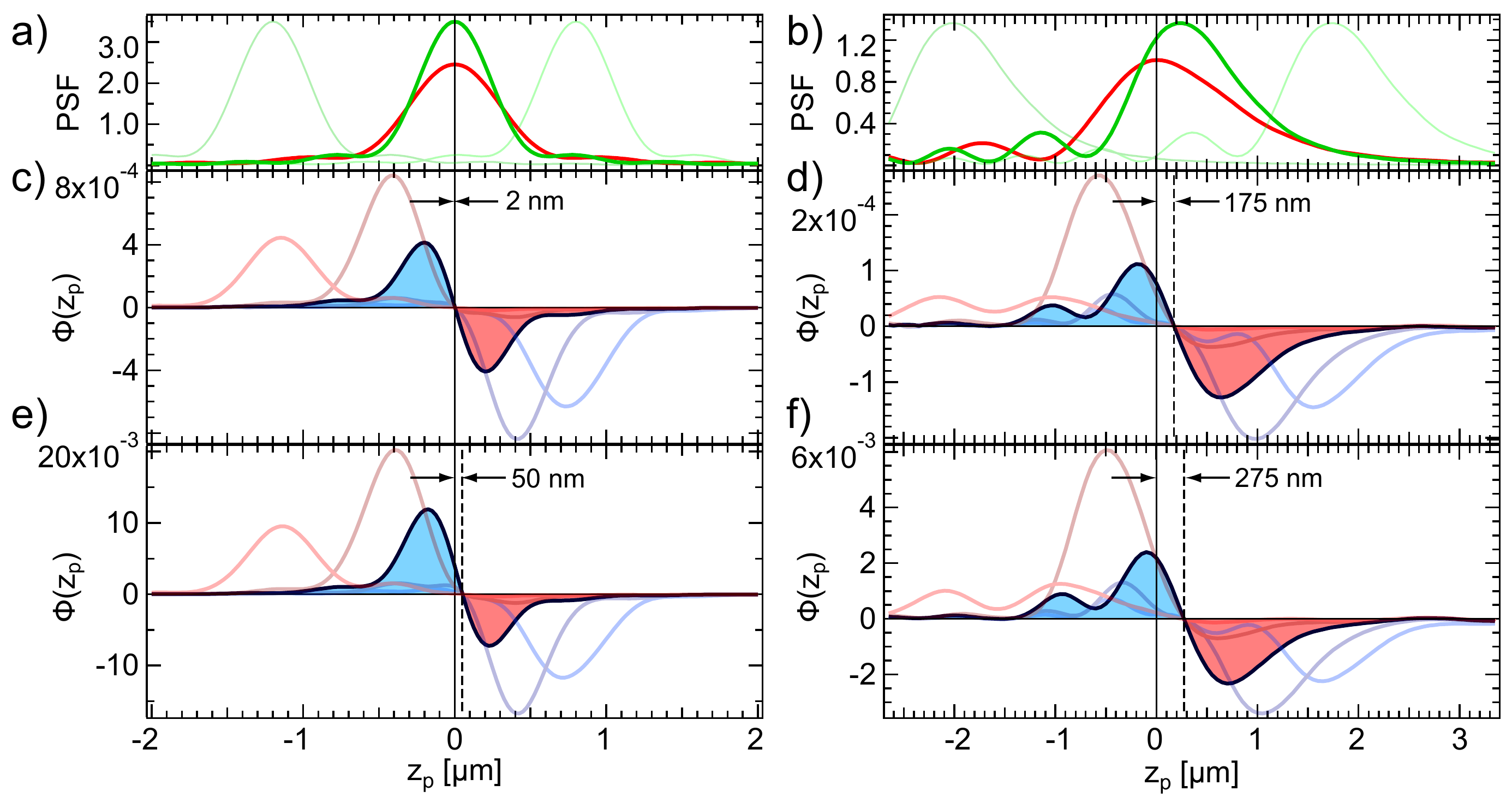}}
\caption{\textbf{a)} Axial scans of the rel.\ PT signal $\Phi\left(z_p\right)$ for $R=\left\{10\,{\rm nm},30\,{\rm nm}\right\}$ (top row, bottom row) AuNPs in PDMS for a heating laser power (peak-to-peak) of $P_{\rm PM}=100\,\mu{\rm W}$ without (left column) and with aberrations (right column). The corresponding incidence beam intensity point spread functions $|\mathbf{E^i}|^2\left(z\right)$ are shown on top. Semi-transpareent curves show the situation of different laser offsets $\Delta z_f$. For details see \cite{SelmkeThesis2013}.}\label{Fig:Aberration}
\end{figure}

\end{widetext}

The most accurate description of the PT signal can be achieved by considering the exact solution to Maxwell's equations for a focused beam scattering. The thermal lens itself can be treated as a (finely) multilayered scatterer, including the heated nanoparticle in its center. As such, the model represents a quantitative description and readily includes finite particle-size effects, the frequency-dependent full thermal lens, aberration effects, pupil-filter effects \cite{Miyazaki2014}, and lateral and axial foci offset and dependencies on the collection angle. The GLMT provides the electromagnetic fields and total cross-section. To evaluate the transmission signal in PT microscopy, it is necessary to consider fractional cross-section. To find these, the time-average of the radial component of the total field's Poynting vector $\mathbf{S^t}=\mathbf{E^t}\times \mathbf{H^t}$ of the total electromagnetic field $\mathbf{E^t}=\mathbf{E^{i}}+\mathbf{E^s}$ is thus evaluated in the far-field ($kr\gg 1$) to find the detectable power contained within a polar angle domain $\left[\theta_{\rm min},\theta_{\rm max}\right]$:
\begin{equation}
P_d\left(\mathbf{r}_p\right)=\lim_{r\rightarrow \infty }\int_{0}^{2\pi}\!\!\int_{\theta_{\rm min}}^{\theta_{\rm max}} r^2 \,\mathbf{\hat{e}_r} \cdot \mathbf{S^t}\left(r,\phi,\theta\right) \mathrm{d}\Omega,\label{eq:PdAngularlyResolved}
\end{equation}
%
%\begin{figure}[bth]
%\centerline{\includegraphics [width=\columnwidth]{Fig_GLMTSchematic.pdf}}
%\caption{Schematic of the discretisation of a thermal lens into a finely multilayered Mie scatterer.}\label{Fig:Frequency}
%\end{figure}
%
In the language of GLMT theory, the scatterer is placed at the origin while the beam is offset by $-\mathbf{r}_p$. The above expression therefore implicitly depends on the particle coordinate $\mathbf{r}_p$. Analogous to the Lorenz-Mie theory, the power is decomposed into three constituents according to $P_d=P_{\rm inc}+P_{\rm sca}+P_{\rm ext}=I_0\left[\sigma_{\rm inc}+\sigma_{\rm sca}-\sigma_{\rm ext}\right]$, i.e.\ incidence, scattering and extinction, respectively. Assuming a full-solid domain, eq.\ \eqref{eq:PdAngularlyResolved} can be rephrased as an equation embodying energy conservation, i.e. $-\sigma_{\rm abs}^\pi=\sigma_{\rm sca}^\pi-\sigma_{\rm ext}^\pi$. Herein, the usual total cross-sections appear and the flux of the incidence beam cancels out between the forwards- and backwards direction. The absorbed power, 
\begin{equation}
P_{\rm abs}=I_0\sigma_{\rm abs}^\pi,
\end{equation}
and thereby the locally induced refractive index perturbation $\Delta n$, is thus conveniently included in such a computation. The introduced fractional cross-sections can be found in Refs.\ \cite{NanoLensDiff,SelmkeACSNano,SelmkeTransmission2014}. As the GLMT is capable of including non-axially positioned probe and pump beams as well, the axial resolution of the PT signal may be studied within this model, beyond the weak scalar diffraction limit discussed in \cite{Cognet2003}. Aberrated beams and finite particle sizes may also be considered such that an exact signal quantification is reachable even under realistic experimental conditions using high numerical aperture objectives, see Fig.\ \ref{Fig:Aberration}a-f). Certain details of the signal which are connected to the finite size of the central particle can only be modelled with this approach as it is currently the only model which includes both the central particle and the thermal lens. For instance, a finite contribution of a modulated scatterer causes a small offset of the zero-crossing of the PT signal even for ideal symmetric beams, see Fig.\ \ref{Fig:Aberration}e,f).

For moderate frequencies with $R_{\rm th}/\omega_0 \gg 1$ it suffices to assume a modulated ideal thermal lens $n\left(r\right)$ and therefore to evaluate the PT signal as the difference between the transmission signal of a bare nanoparticle and a nanoparticle surrounded by a thermal lens. 
\begin{equation}
\Phi=\frac{\left(\sigma_{\rm sca} - \sigma_{\rm ext}\right)_{{\rm AuNP} + n\left(r\right)}-\left(\sigma_{\rm sca} - \sigma_{\rm ext}\right)_{\rm AuNP}}{\sigma_{\rm inc}}\label{eq:GLMTPhi}
\end{equation}
A two-lobed axial structure $\Phi\left(z_p\right)$ characterizes the PT signal shape. While present already in early PT experiments \cite{Boyer2002,Paulo2009}, a detailed study was first done in Refs.\ \cite{SelmkeACSNano,SelmkeABCD,SelmkeThesis2013}. The axial signal dependence mirrors that of a probing beam which passes a lens and is collected using an aperture. This indeed corresponds to the situation in PT detection in transmission. In addition to the simple notion of a product of the focal intensity distributions of both lasers, the signal shape is thus found to possess a zero-crossing around the particle location as a result of the lensing. As a further consequence of the thermal lens acting as a large transparent Mie scatterer, the forward lensing signal increases linearly with the refractive contrast $\sigma_{\rm sca}\left(\Delta n\right)-\sigma_{\rm ext}\left(\Delta n\right)\propto \Delta n$ although its individual constituents $\sigma_{\rm sca, ext}\left(\Delta n\right)$ vary nonlinearly. The significance of the scattering contribution and the non-linearity grows for increasing collection angles. It is only the total change in transmittance due to scattering and interference together, which is linear in the perturbation. The same can be found for transparent microspheres which act as lenses on a probing beam and represent the limiting case of macroscopic lenses.

If the frequency-dependence shall be considered explicitly, the full time-like evolution of the detected power $P_d\left(t\right)$ must be evaluated over a time-interval $T=2\pi/\Omega$ corresponding to the period of heating. The relative transmission $P_d\left(t\right)/P_{\rm inc}$ accordingly varies over one cycle. The corresponding in-phase rel.\ PT signal $\Phi_{\cos}$ and the out-of-phase rel.\ PT signal $\Phi_{\sin}$ may then be obtained via 
\begin{align}
\Phi_{\cos}=& \frac{1}{T}\int_{0}^{T}\cos\left(\Omega t\right)\frac{P_d\left(t\right)}{P_{\rm inc}}\mathrm{d}t,\label{eq:PhiCos}\\
\Phi_{\sin}=& \frac{1}{T}\int_{0}^{T}\sin\left(\Omega t\right)\frac{P_d\left(t\right)}{P_{\rm inc}}\mathrm{d}t.\label{eq:PhiSin}
\end{align}
The result of such a computation provides a quite different picture than the problematic plane-wave treatment in the Born approximation. While both predict a decay of the absolute value of the PT signal according to $kR_{\rm th}^2$, the composition is quite different. Also, only in the GLMT framework a saturation of the signal is found for $kR_{\rm th}\rightarrow \infty$, due to the finite probe beam size. Further, for very short-ranged thermal lenses with $kR_{\rm th}\ll 1$, only the GLMT reveals a saturation at an in-phase signal which accounts for a modulated particle scatterer in an effectively modulated environment refractive index. Both predictions and their comparison the the absolute value are shown in Fig.\ \ref{Fig:Frequency}.

The resolution of the PT microscopy technique may be estimated by its detection point-spread function, that is, via the characteristic widths of a signal scan $\Phi\left(x_p,z_p\right)$ across a point-like absorber. The GLMT allows to predict this quantity and may also be used for pupil-filtered illumination. Recently, a resolution enhancement was found using an annular pupil filter \cite{Miyazaki2014}.

%\begin{widetext}
%
%\begin{figure}[bt]
%\centerline{\includegraphics [width=\columnwidth]{Fig_GLMTCombined.pdf}}
%\caption{\textbf{a)} Axial scans of the rel.\ PT signal $\Phi\left(z_p\right)$ for $R=\left\{10\,{\rm nm},30\,{\rm nm}\right\}$ (top row, bottom row) AuNPs in PDMS for a heating laser power (peak-to-peak) of $P_{\rm PM}=100\,\mu{\rm W}$ without (left column) and with aberrations (right column). The corresponding incidence beam intensity point spread functions $|\mathbf{E^i}|^2\left(z\right)$ are shown on top. Semi-transpareent curves show the situation of different laser offsets $\Delta z_f$. For details see \cite{SelmkeThesis2013}. \textbf{g)} Screening length dependence of the rel.\ PT signal, $\Phi\left(R_{\rm th}\right)$ according to the accurate GLMT model ($R=10\,\rm nm$, thick lines: $\omega_0=281\,\rm nm$ and ${\rm NA}_d=0.8$ thin lines $\omega_0=500\,\rm nm$ and ${\rm NA}_d=0.5$) and \textbf{h)} the plane-wave Born approximation theory. \textbf{i)} Axial scans of the in-phase $\Phi_{\cos}\left(z_p\right)$ and out-of-phase $\Phi_{\sin}\left(z_p\right)$ contributions for various scaled thermal diffusion lengths $k R_{\rm th}$.}\label{Fig:GLMTCombined}
%\end{figure}
%
%\end{widetext}

%

%\begin{figure}[bth]
%\centerline{\includegraphics [width=\textwidth]{Fig_GLMT.pdf}}
%\caption{\textbf{Top group:} Polarization resolved transmission scans $P_d\left(x_p,z_p\right)$ under realistic illumination conditions using high NA microscope objectives (${\rm NA}_{\rm ill}=1.4$, ${\rm NA}_d=0.8$). Probe laser ($\lambda=635\,\rm nm$, $\omega_0=280\,\rm nm$) and a pump laser ($\lambda=532\,\rm nm$, $\omega_0=230\,\rm nm$, $P_h=100\,\rm mW$ peak-to-peak, $\Omega=300\,\rm kHz$) illuminate a gold nanoparticle ($R=30\,\rm nm$) in PDMS ($n_0=1.46$). \textbf{Bottom row:} Resulting relative PT signal $\Phi\left(z_p\right)$ for three relative laser configurations, $\Delta z_f\approx \left\{-z_R,0,+z_R\right\}$.}\label{Fig:Rutherford}
%\end{figure}

\section{Photonic Rutherford scattering (D)\label{sec:photRF}}
In this model the photothermal effect on the propagation of the probe beam is framed in an analogy to Rutherford scattering. It formalizes the apparent similarity which may be anticipated from the functional dependence of the refractive index profile of the ideal thermal lens, where $\Delta n\left(r\right)\propto 1/r$. As such, the framework naturally exposes a deflection phenomenon which is intrinsic to any PT measurement. The model also provides a near field picture of the thermal lensing effect on the focused probe beam and may be applied to give quantitative results as well. The Helmholtz equation \eqref{eq:HE} may be solved analytically \cite{SelmkeAmJPhys2013,SelmkePRL2013} for the scattering potential $F$ to linear order in the perturbation $\propto \nu=-k \Delta n R /n_0$ and the ideal thermal lens eq.\ \eqref{eq:nrIdeal}.
\begin{equation}\label{eq:HETL}
\nabla^2 U_C^k+k^2\left[1-\frac{2 \nu}{ k r}\right]U_C^k=0=0. %+\left\{\frac{\Delta n^2 R^2}{n_0^2 r^2}\right\}
\end{equation}
The solution to the resulting differential equation under the assumption of an incident plane wave was given by W.\ Gordon in the context of the wave-mechanical description of Rutherford scattering,
\begin{equation}
\frac{U^{\mathbf{k}}_C\left(\mathbf{r}\right)}{U_0}=e^{-\frac{\pi}{2}\nu}e^{i\mathbf{k}\cdot\mathbf{r}}\Gamma\left(1+i\nu\right) {}_{1}F_{1}\left(-i\nu;1;i\left(kr - \mathbf{k} \cdot \mathbf{r}\right)\right),\label{SchroediExactSolution}
\end{equation}
with $_1 F_1$ and $\Gamma$ denoting the confluent hypergeometric function and the gamma function, respectively. Again, the solution represents the total field including the incidence and the scattered field. In the forward direction the amplitude is finite and approximately equal to $|U^{\mathbf{k}}_C|^2/|U_0|^2 \approx 1-\pi\nu$. Rather than a diverging forward scattering amplitude as for plane-wave scattering, the forward amplitude disturbance is $-\pi \nu$. As we will see, it is this forward interference which determines the PT signal even in case of a focused beam, where it is smeared out to the angular domain of propagation determined by the beam's angle of divergence $\theta_{\rm div}=2/k\omega_0$. Thus far, the signal appears linear in $\nu$ instead of the refractive index contrast. Again, the correct dependence will emerge only after the inclusion of a focused beam probing. Here, a focused beam may be implemented by assuming an incident wave-packet $U_0^{\rm wp}\left(\mathbf{r}\right)=U_0\int\mathrm{d}\mathbf{k}\, A\left(\mathbf{k}\right) e^{i \mathbf{k}\cdot\left(\mathbf{r}-\mathbf{r_0}\right)}$ with an azimuthally symmetric monochromatic wave-vector spectrum $A\left(\vartheta\right)=\delta\left(\bar{k}-k\right)\exp\left(-\vartheta^2/2\sigma_\vartheta^2\right)$ in spherical coordinates. Such a wave-packet resembles the TEM-00 mode Gaussian beam and has a characteristic width-scale given by $\omega_\vartheta=2/ \left[k\sigma_\vartheta\right]$. The modified probing beam is given by the superposition of the plane-wave solutions eq.\ \eqref{SchroediExactSolution}:
\begin{equation}
U^{\rm wp}_C\left(\mathrm{r}\right)=\int\!\mathrm{d}\mathbf{k}\,A\left(\mathbf{k}\right) e^{-i \mathbf{k}\cdot\mathbf{r_0}}\, U^{\mathbf{k}}_C\left(\mathbf{r}\right). \label{eq:TDSEgeneral}
\end{equation}
Using the above formalism the near-field structure of the photothermal signal becomes accessible and the characteristic energy redistribution close to the particle may be seen to be responsible for the far-field signal. A quantitative analysis of the far field PT signal can be done by considering integrated differences in relative changes of intensities, i.e.\ in analogy to the evaluation of the signal in the diffraction framework, eq.\ \eqref{eqPhiApproxDiffrOnAxis}. The results agree in the paraxial limit corresponding to moderate focusing and are in concord with the rigorous GLMT model as well, see the supplement of Ref.\ \cite{SelmkePRL2013}.

\begin{figure}[tb]
\centerline{\includegraphics [width=\columnwidth]{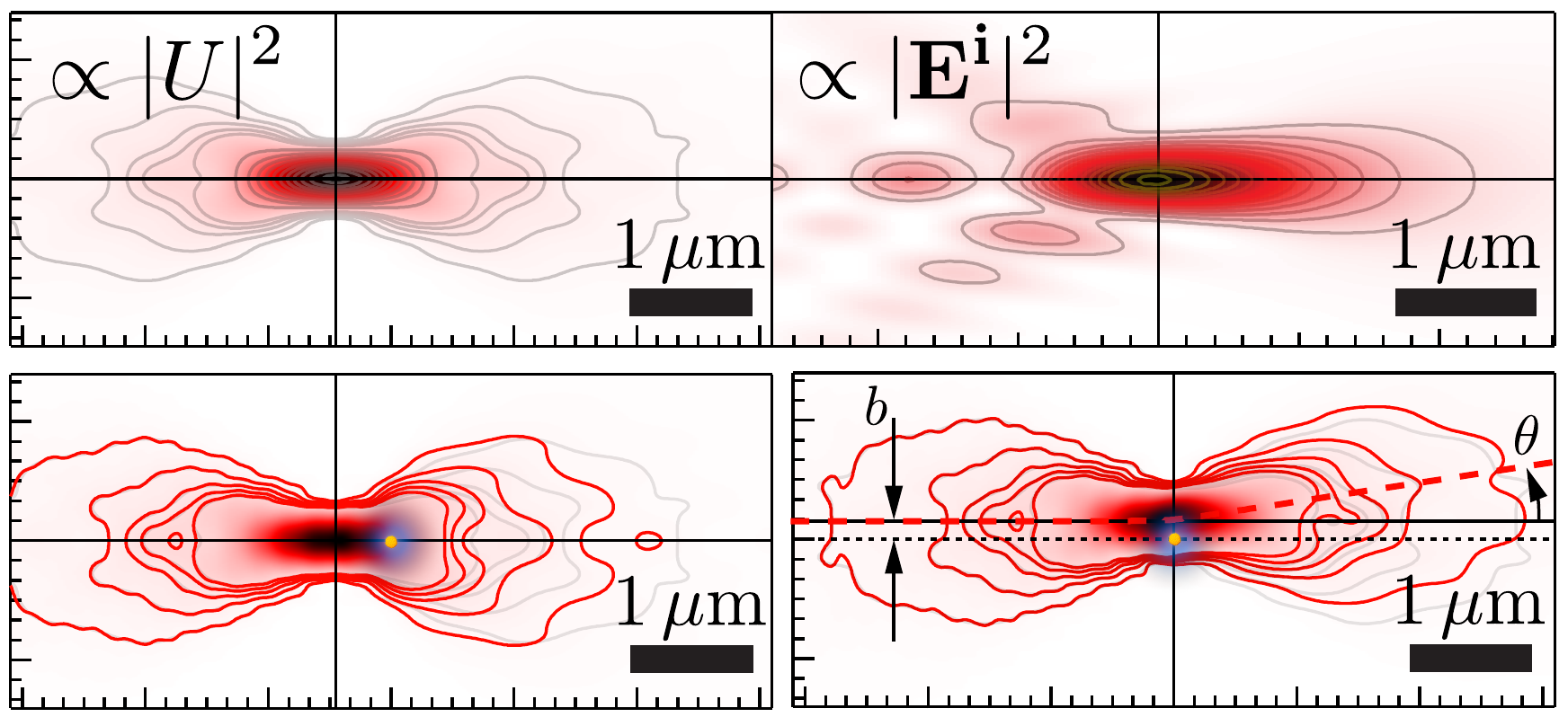}}
\caption{Near field probe beam intensity. The top row shows an ideal Gaussian wave-packet $|U^{\rm wp}|^2$, eq.\ \eqref{eq:TDSEgeneral}, and a realistic aberrated focal intensity distribution $|\mathbf{E^i}|^2$. The bottom row shows a lensing and a deflection scenario. Gray solid lines: Contours of the incidence field $|U^{\rm wp}|^2$. Red lines: Contours of the affected probe beam $|U^{\rm wp}_C|^2$. The red dashed line shows the ray-trajectory through the TL, i.e.\ Eq.\ \eqref{eq:Rutherford_Photon}.}\label{Fig:Rutherford}
\end{figure}

The effect of finite modulation frequencies beyond the quasi-static limit can be incorporated in a similar manner. Using the plane-wave scattering results on the full thermal wave, Eq.\ \eqref{eq:fomega}, a plane-wave decomposition may be used to construct the harmonic contributions to the PT signal for any shaped / focused beam along the lines shown above. Using this approach, the optimal detection numerical aperture can be determined at any frequency \cite{Miyazaki2014Optimal}. As a consequence of the energy-redistribution \cite{SelmkeTransmission2014}, the detection aperture should always be taken to be smaller than the illumination aperture \cite{NanoLensDiff,Miyazaki2014Optimal} such that the lensing signal is efficiently detected and the signal-to-noise ratio is optimal.

\paragraph{Classical limit and beam-centroid deflection}
It was shown that the wave packet which represents the focused beam exactly follows the classical Rutherford trajectories in the proper limit. This classical limit of geometrical Rutherford scattering corresponds to the parametric ray solution to Fermat's least optical path principle 
\begin{equation}
\frac{\mathrm{d}^2\mathbf{r}}{\mathrm{d}s^2}=\mathbf{\nabla}\frac{1}{2}n^2\left(\mathbf{r}\right), \quad \left|\frac{\mathrm{d}\mathbf{r}}{\mathrm{d}s}\right|=n\left(\mathbf{r}\right)\label{eq:Fermat}
\end{equation}
with the stepping parameter $s$ along the ray. For weak and ideal thermal lenses of the form \eqref{eq:nrIdeal}, hyperbolic ray trajectories $r\left(\phi\right)$ solve the above differential equation:
\begin{equation}\label{eq:Rutherford_Photon}
r\left(\phi\right)\approx \frac{|\xi| b^2}{\sqrt{b^2\xi^2+1}\cos\left(\phi-\phi_0\right) \pm 1}
\end{equation}
in polar coordinates. Herein, the appropriate sign $\pm 1=-\xi/|\xi|$ is determined by sign of the interaction strength in $\xi=k/\nu=-n_0/\Delta n R$, which appears here in a wavelength independent combination and is taken to be weak ($b\xi\ll 1$). The full solution contains a further perturbation parameter, see \cite{SelmkeAmJPhys2013}. The ray was assumed incident with a distance to the optical axis of $b$. A corresponding deflection by an angle $\theta_d=2\phi_0-\pi$ determined by
\begin{equation}
\cot\left(\theta_d/2\right)=b\xi,\label{eq:costhetahalfRF}
\end{equation}
occurs, even for a highly focused beam for off-axis illumination under typical experimental conditions, as shown in Ref.\ \cite{SelmkePRL2013}. However, unless purposefully detected by a position sensitive detector, predominantly the lensing will be responsible for the transmission change and thereby the PT signal. In general, however, referring to the classical spectroscopy variants, one may expect both the lensing and the deflection signal to be of the same order of magnitude \cite{Jackson1981}. The ray description proves to be particularly useful when combined with the methods of Gaussian beam transformation optics, see Sec.\ \ref{sec:ABCD}.

\section{Fresnel diffraction / PIC (E)\label{FresnelSec}}
In photothermal interference contrast microscopy it is directly the phase-shift which is used for imaging of absorbers \cite{Boyer2002}, see Fig.\ \ref{Fig:Schemes}e). A first investigation of the induced phase-shift due to a thermal lens was done in Ref.\ \cite{Cognet2003} in the thin phase-grating approximation. The time-dependent thermal lens $n\left(r,t\right)$ may then be approximated as a sphere of constant index of refraction, whose size is estimated via $\rho_{\rm th}\sim R_{\rm th}$. The probing beam has to propagate through this region in which the averaged refractive index change is taken to be constant at $\overline{\Delta n}\sim \Delta n$. Together, these estimations are used to provide a crude approximation for the accumulated phase-shift through this hypothetical phase-sphere, $\Delta \chi \sim k \rho_{\rm th} \overline{\Delta n}$, which is maximum if the particle is in the focus. These considerations already show that the induced refractive index change is small and typically in the order of some $\overline{\Delta \chi} \sim 10^{-4}\, \rm rad$ \cite{Cognet2003}.

Elaborating on this notion of the thermal lens as a phase-modifying element, a quantitative model was formulated for the PT signal in the direct transmission detection scheme, see Fig.\ \ref{Fig:Schemes}d). While the following discussion is thus tailored towards the single beam probing and its self-interference, the method is likely to be applicable in the PIC detection scheme as well. The thermal lens will now be taken to be the ideal TL $n\left(r\right)$ which is of infinite spatial extent. An analytic solution to the PT signal can then be found in the paraxial limit of a weakly focused probing field \cite{NanoLensDiff,SelmkePRL2013}. Thereby, the signal magnitude along with its axial position and angular dependence will be accessible. The approximation involves the assumption of a Gaussian beam and the Fresnel-diffraction integral, which both represent a solution to the paraxial scalar Helmholtz equation. In the language of diffraction, the TL then acts as a variable phase mask. The scalar field amplitude $U\left(x,z\right)$ in the image-plane located at a distance $z$ behind the aperture-plane is given by the following diffraction integral:
\begin{equation}
U=\frac{k}{iz}\, e^{ikz+i\frac{kx^2}{2z}}\!\! \int_{0}^{\infty} \!\!\! \! U_a\, e^{i \frac{k\rho^2}{2z}} J_0\left(\frac{k\rho x}{z}\right)\mathcal{A}\left(\rho\right)\rho\,\mathrm{d}\rho.\label{DiffractionFocused}
\end{equation}
Herein, $U_a\left(\rho,z_p\right)$ is the beam field in the aperture-plane at $z=0$, centred arbitrarily at the position of the particle (therefore depending on the particle coordinate $z_p$). The function $\mathcal{A}=\exp\left(i \Delta \chi_\nu\right)$ represents the radially symmetric variable phase mask which encodes the TL placed on the optical axis. The solution, $U$, may be identified with the $x$-component of the total field amplitude $\mathbf{E^i}+\mathbf{E^s}$. It therefore embodies the incidence, scattering and interference contributions at the same time. In a straight-ray approximation, the resulting collected phase advance $\Delta \chi_\nu$ for the ideal thermal lens yields
\begin{equation}
\Delta \chi_\nu \left(\rho\right) =k_0\!\int_{-L}^{L} \!n\left(\!\sqrt{z^2+ \rho^2}\right) \mathrm{d} z \approx 2\nu \ln\left(\rho\right)\nonumber \label{EqPhaseAdvance},
\end{equation}
where additional constant phases were discarded. The travelled distance $L$ in front and behind the TL was assumed to be large as compared to the contributing radii $\rho$ in the integration. This is ensured by the exponential decay of the illuminating aperture field $U_a$. However, as it turns out, this is not even necessary. In fact, considering a plane-wave illumination in the diffraction integral, $U_a\rightarrow 1$, the exact solution $U_C^{k\mathbf{\hat{z}}}$, eq.\ \eqref{SchroediExactSolution}, to the full Helmholtz equation as encountered in photonic Rutherford scattering, is recovered up to a logarithmic phase-factor \cite{SelmkePRL2013}. Also, the thermal lens is seen to exhibit a special kind of radial dependence which is quite different from a spherical lens. This quantifies the spherical aberrations of the ideal TL, see Table \ref{table:Transmissionfunctions}. In the case of a Gaussian beam we can now write for the rel.\ PT signal $\Phi$ on the $z$-axis, using Eq.\ \eqref{DiffractionFocused}:
\begin{equation}
\Phi = \frac{|U_\nu|^2 - |U_{\nu=0}|^2}{|U_{\nu=0}|^2} \approx 2\nu\left(z_p\right) \arctan\left(\frac{-z_p}{z_R}\right)\label{eqPhiApproxDiffrOnAxis},
\end{equation}
to first order in $\mathcal{O}\left(\nu\right)$. The result reduces to the limiting value $\Phi\rightarrow -\pi\nu$ for plane-wave incidence, which coincides with the exact forward amplitude of an incident plane wave. Note that $\arctan()$ is \textit{not} resulting from the beam's Gouy-phase, which cancels out in Eq.\ \eqref{eqPhiApproxDiffrOnAxis}. Instead, it is the result of the specific phase-shift for the long-ranged TL profile. Again, for some fixed small offset $z_p\ne 0$ the signal is non-zero and proportional to $\Phi \propto \Delta n /z_R \propto \Delta n/n_0$, i.e.\ the optical contrast.

\begin{table}[bt]
\begin{center}
\begin{tabular}{|l|c|}
\hline
kind of lens & \hspace{1.4cm} phase $\Delta \chi$ \hspace{1.4cm} \\ \hline
thermal lens & $\displaystyle 2k_0\left[L n_0 + R\,\Delta n \ln\left(\frac{L+\sqrt{L^2+\rho^2}}{\rho}\right)\right]$ \\
thick lens $d\left(\rho\right)$ & $\displaystyle k_0 \left[n_0 d_0 + \Delta n_L d\left(\rho\right)\right]$ \\
thin spherical lens & $\displaystyle -k_0 n_0 \frac{\rho^2}{2f}$\\
\hline
\end{tabular}
\caption[Transmission functions of the thermal and a thin lens]{Transmission functions for the thermal lens and for lenses of constant index of refraction.\label{table:Transmissionfunctions}}
\end{center}
\end{table}

\paragraph{The angular signature of the PT signal}
Using the Fresnel diffraction integral also the angular pattern $\Phi\left(\theta,z_p\right)$ of the PT signal for on-axis probing may be found \cite{NanoLensDiff}. For a probing beam positioned behind the lens ($z_p <0$), the angular pattern shows a peak towards the center and an annular dip at larger angles. For the case of a probing beam being positioned in front of the lens ($z_p >0$), the angular pattern simply changes its sign, relative to the previous scenario. These principal features resemble those which are found in thermal lens spectroscopy of macroscopic samples, although the thermal lens geometry is then cylindrical. Overall, the energy of the probe beam is redistributed by the action of the lens, i.e.\ an integration of $\Phi\left(\theta,z_p\right)\sin\left(\theta\right)$ from $0$ to $\pi/2$ gives zero. The pattern of this energy redistributions is consistent with the to the notion of a broadened beam, whereby the angular pattern would be the difference of two Gaussians, as put forward in the ABCD model in section \ref{sec:ABCD}.

From the angular pattern one can determine the total detected photothermal signal for a finite angular detection domain via an integration,
\begin{equation}
\Phi=\frac{4\theta_{\rm div}^{-2}}{e^{-\vartheta_r\left(\theta_{\rm min}\right)}-e^{-\vartheta_r\left(\theta_{\rm max}\right)}}\int_{\theta_{\rm min}}^{\theta_{\rm max}}\Phi\left(\theta,z_p\right)\frac{\sin\left(\theta\right)}{\cos^{3}\left(\theta\right)}\mathrm{d}\theta ,\label{EqSignalNAd}
\end{equation}
\begin{figure}[b]
\centering
	\includegraphics [width=\columnwidth]{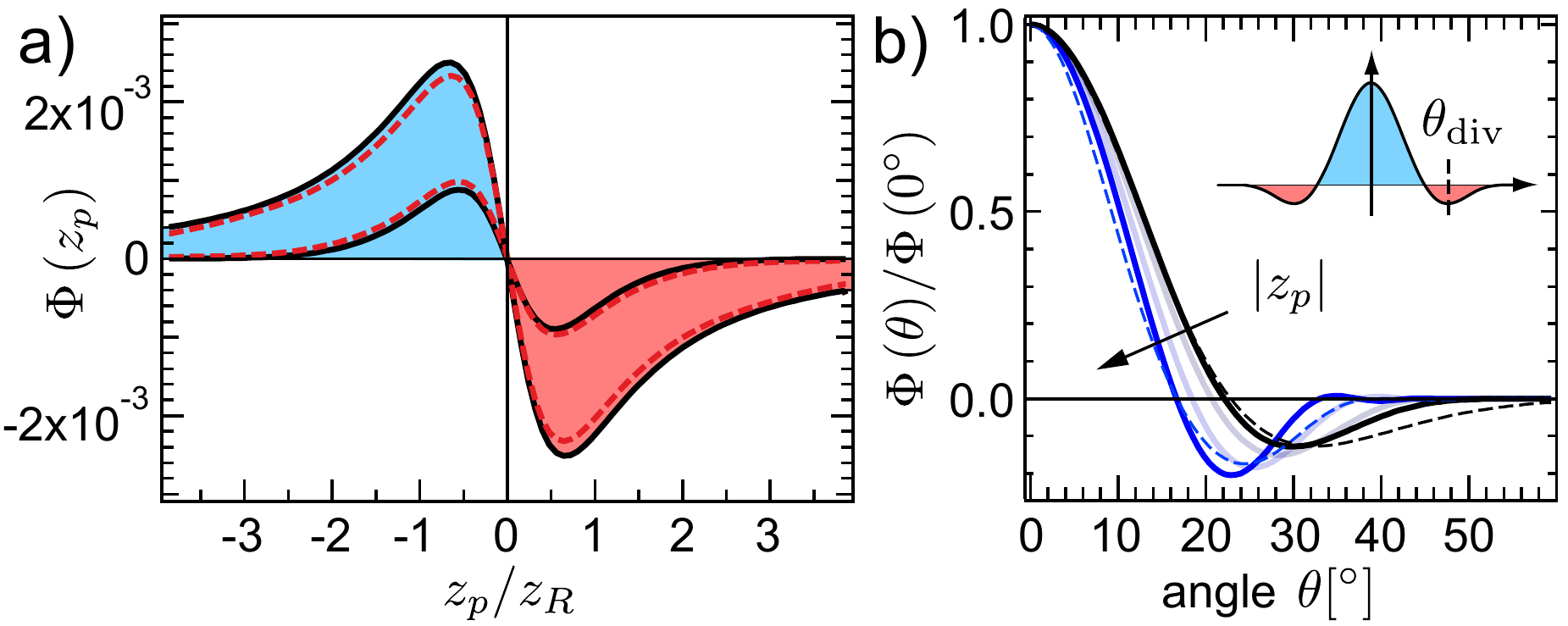}
\caption{\textbf{a)} On-axis $z_p$-scan for $\mathrm{NA}_d=0$ (larger signal) and ${\rm NA}_d=0.75$ of the rel.\ PT signal $\Phi$. Solid and dashed lines correspond to the diffraction model Eq.\ \eqref{EqSignalNAd} and the GLMT model Eq.\ \eqref{eq:GLMTPhi}, respectively. \textbf{b)} Angular spectrum of the PT signal for a divergence angle $\theta_{\rm div}\approx 30^{\circ}$ and for $z_p=\left\{0.01z_R,z_R,2z_R,3z_R\right\}$, normalized to the forward direction. \label{Fig:DiffrGLMTCompare}}
\end{figure}
which again, as in section \ref{sec:ABCD}, depends on the beam's angle of divergence through the reduced quantity $\vartheta_r\left(\theta\right)=2\tan^2\left(\theta\right)/\theta_{\rm div}^2$. Both, the angular pattern of the PT signal and consequently its integrated value for finite detection angles agree with those found using the rigorous GLMT formalism (see Fig.\ \ref{Fig:DiffrGLMTCompare}) .

\begin{figure}[tbh]
\centering
	\includegraphics [width=\columnwidth]{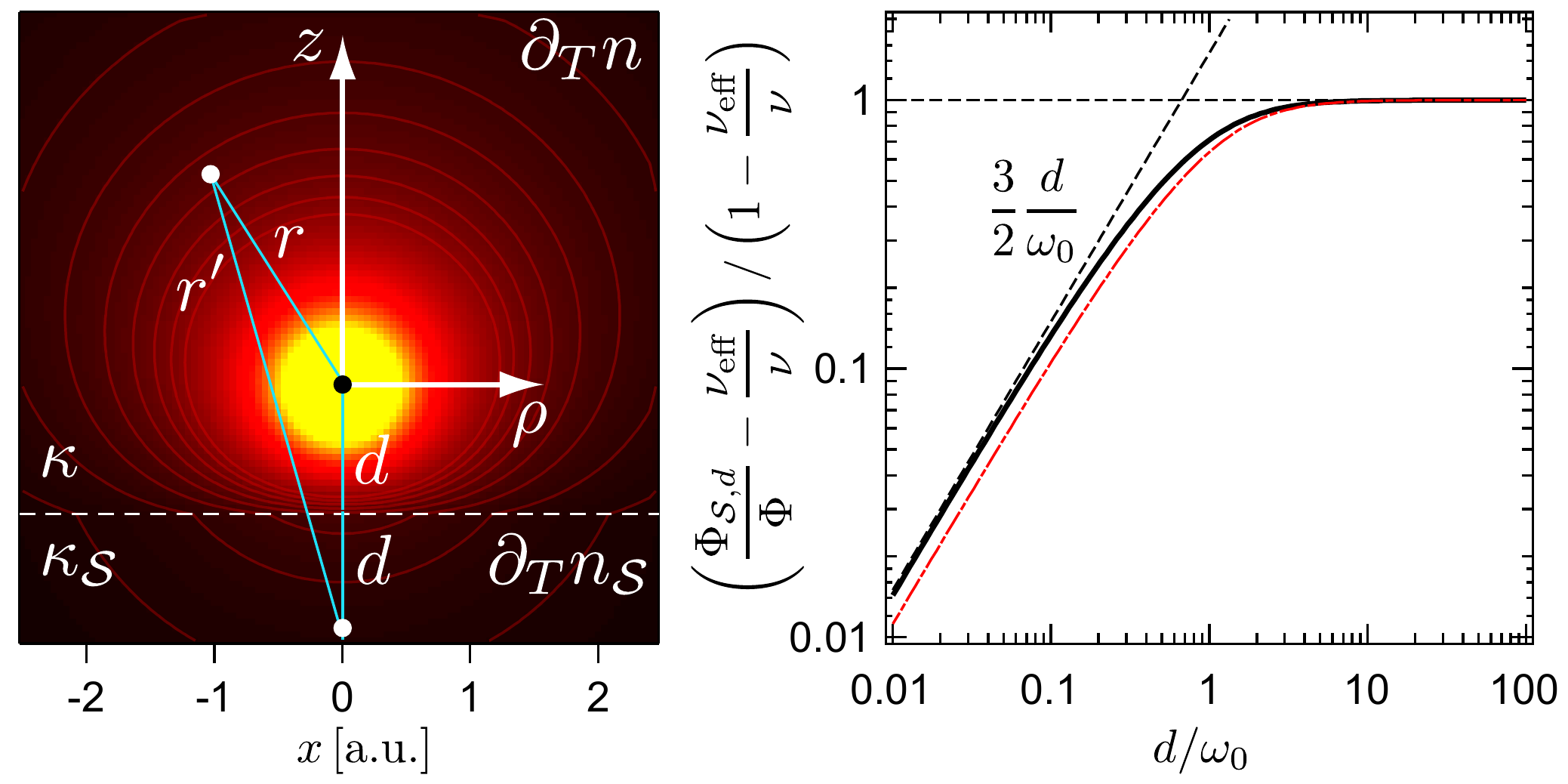}
\caption{\textbf{a)} Rel.\ PT signals for thermal lenses with $\nu=10^{-3}$ above a half-space with $\partial_T n_{\mathcal{S}} = \partial_T n$. The heat-source is embedded in a material with $n_0=1.46$, $\kappa=0.15\mathrm{W}/\mathrm{K m}$ and close to an interface. Signal at $z_p=-z_R$ (black) and $-2z_R$ (red, dashed-solid) as a function of the heat-source distance $d/z_R$ for $\kappa_{\mathcal{S}}=\left\{1,10\right\}\mathrm{W}/\mathrm{K m}$ (dashed, dotted black lines).\label{Fig:Interface}}
\end{figure}

\paragraph{Interface effects}
The Diffraction method is well-suited to analyse the effects due to a plane interface near the single absorber. To this end, one may use the solution to the heat equation for a point-like heat source placed at a hight $d$ above an interface \cite{Yovanovich1995}. As it is based on the method of images, the result is framed as the sum of point-sources, allowing analytic progress in analogy the the previous discussion.
\begin{equation}
n\left(\mathbf{r}\right)=\left\{ \!\! \begin{array}{c} n_0 \\ n_{0,\mathcal{S}} \end{array}\!\! \right\}  + \Delta n\left\{ \begin{array}{c}
\displaystyle\frac{R}{r}+\left(\frac{\kappa-\kappa_{\mathcal{S}}}{\kappa+\kappa_{\mathcal{S}}}\right)\frac{R}{r'}\\
\displaystyle N \!\left(\frac{2\kappa}{\kappa+\kappa_{\mathcal{S}}}\right)\frac{R}{r}\end{array} 
\right. \label{eqThermalLens},
\end{equation}
where the upper und lower lines corresponds to the two half-spaces with $z\ge 0$ and $z<0$, respectively. Here, the refractive index contrast $\Delta n = \left(\partial_T n\right) P_{\rm abs}/4\pi \kappa R$ assumes a homogeneous medium, and $N=\partial_T n_{\mathcal{S}}/\partial_T n$  is the ratio of the thermorefractive coefficients of the support material and the embedding medium. Further, $r^2=\rho^2+\left(z-d\right)^2$ is the squared distance of point $\mathbf{r}=\left(\rho,z\right)$ to the actual heat source placed at $\left(0,d\right)$, and $r'^2=\rho^2+\left(z+d\right)^2$ is the squared distance of the same point to the image heat source situated at $\left(0,-d\right)$, see Fig.\ \ref{Fig:Interface}. The computation of the signal via the diffraction integral eq.\ \eqref{DiffractionFocused} requires the total phase advance $\Delta \chi_\nu=\Delta \chi_1+\Delta \chi_2+\Delta \chi_3$, which is now the sum of three terms each computed analogously to eq.\ \eqref{EqPhaseAdvance} with the proper limits.

The accumulated phase in the straight-ray approximation, again up to constant terms, amounts to
\begin{equation}
\Delta \chi_\nu \left(\rho\right)= 2\nu_{\rm eff}\ln\left(\rho\right) +2\nu' \ln\left(\frac{d}{\rho}+\sqrt{1+\frac{d^2}{\rho^2}}\right),
\end{equation}
with the rescaled strengths related to $\nu=-k R \Delta n/n_0$
\begin{equation}
\nu_{\rm eff}=\nu \frac{\kappa}{\kappa_{\rm eff}} \frac{\left(1+N\right)}{2}, \quad \nu'=\nu\frac{\left(N\kappa-\kappa_{\mathcal{S}}\right)}{2\kappa_{\rm eff}},
\end{equation}
For $d=0$, $\kappa=\kappa_{\mathcal{S}}$ and $N=1$ the result of the ideal thermal lens is recovered. For $d=0$, $N=1$ but $\kappa\ne \kappa_{\mathcal{S}}$ the effective thermal conductivity $\kappa_{\rm eff}=\left[\kappa + \kappa_{\mathcal{S}}\right]/2$ appears. The expressions for the rel.\ PT signal and its angular spectrum in this case correspond to those of an ideal thermal lens. For $d=\infty$ the interface may be neglected as $\Delta \chi_\nu\rightarrow 2\ln\left(\rho\right)\left[\nu_{\rm eff}-\nu'\right]=2\nu\ln\left(\rho\right)$ and the signal approaches the value corresponding to no interface.

The rel.\ PT signal on the optical axis is then:
\begin{equation}
\Phi_{\mathcal{S},d} = 4|\varsigma|^2 \left|\int_{0}^{\infty} \exp\left(-\varsigma \rho^2\right)\rho \exp\left(i\Delta\chi_\nu\right) \mathrm{d}\rho\,\right|^2 - 1\label{eqPhiApproxDiffrOnAxis},
\end{equation}
with the abbreviation $\varsigma\left(z,z_p\right)=1/\omega^2\left(z_p\right)-i k/\left[2R_C\!\left(z_p\right)\right]$ due to the input Gaussian beam field. A similar expression may be found for finite numerical aperture. For a source situated directly on the interface the expression for the rel.\ PT signal simply becomes
\begin{equation}
\Phi_{\mathcal{S}}\left(z_p\right) \approx 2\nu_{\rm eff}\arctan\left(\frac{-z_p}{z_R}\right) \label{eqPhiApproxDiffrOnAxisd0}
\end{equation}
with the effective strength parameter $\nu_{\rm eff} \propto P_{\rm abs} \left[\partial_T n+\partial_T n_{\mathcal{S}}\right] \kappa_{\rm eff}^{-1}$. The rel.\ PT signal in the general situation is bounded by the extreme values for zero and infinite heat-source distance $d$, i.e.\ the rel. PT signal obeys the inequality $\Phi_{\mathcal{S}}=\Phi \left[\nu_{\rm eff}/\nu\right]< \Phi_{\mathcal{S},d} < \Phi$. The transition between these cases is smooth and naturally scales with the beam-waist $\omega_0$, which can be seen upon inspection of the integrand in eq.\ \eqref{eqPhiApproxDiffrOnAxis}. The transition is shown in Fig.\ \ref{Fig:Interface}. The transition as a function of $d/\omega_0$ is insensitive to the ratios $\kappa/\kappa_{\mathcal{S}}$ or $N$ and only weakly sensitive to a probing offset $|z_p| \sim z_R$.
The analytical diffraction model may also be applied to a situation where an interface of finite thermal interfacial conductance is nearby. The thermal lens may again be expressed by point-like image sources, although continuous in this case \cite{WangInterfaceConductivity}.

\section{Gaussian ABCD matrix transformation (F)\label{sec:ABCD}}
\begin{widetext}

\begin{figure}[t]
\centering
	\includegraphics [width=\columnwidth]{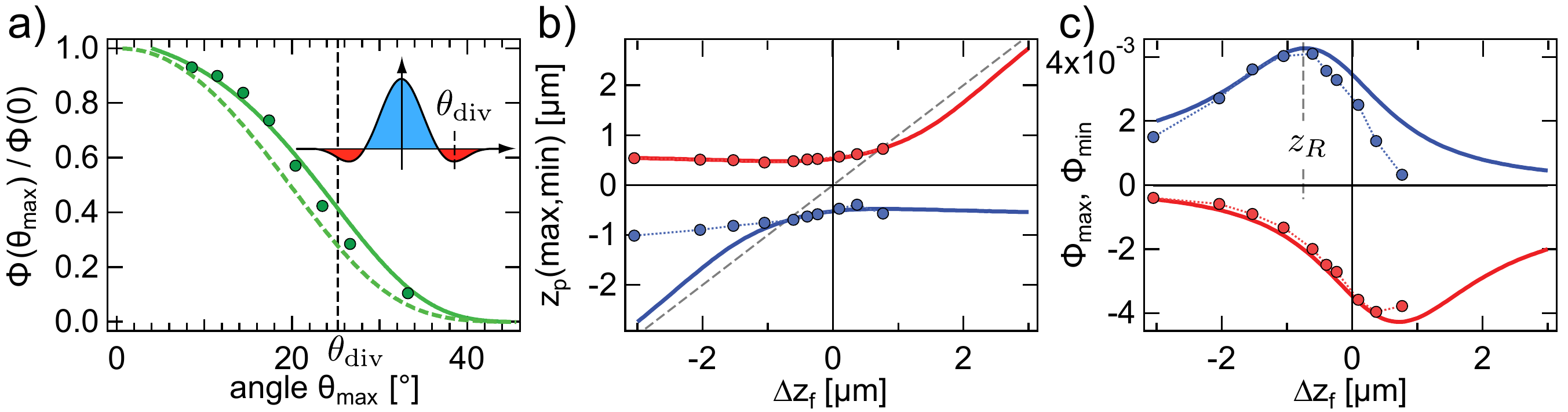}
\caption{\textbf{a)} Normalized $\Phi$-dependence on the collection angle $\theta_{\rm max}$ at $z_p=-0.5\mu{\rm m}$: Experimental (green circles, weakly focused beams), model \textbf{(F)} (solid green line), model \textbf{(E)} (green dashed line). The decrease is due to the angular pattern $\Phi\left(\theta\right)$ schematically depicted as the inset. The PT signal $\propto \Delta P_d$ vanishes when the angular detection domain extends far across the probe beam's angle of divergence $\theta_{\rm div}$. \textbf{b)} Extrema positions vs.\ axial displacement $\Delta z_f$ of heating and probe laser. The \textsc{Rayleigh}-range of the probe beam is $z_R=0.72\,\mu\rm m$. \textbf{c)} Extrema values $\Phi_{\rm max}$ and $\Phi_{\rm min}$ of the rel.\ PT signal.\label{FigGaussABCD}}
\end{figure}

\end{widetext}

A powerful and intuitive analytical model of the PT signal has been put forward using ABCD Gaussian beam transformation optics \cite{SelmkeABCD}. Similar models have been put forward in thin slab thermal lensing spectroscopy. The difference rests in the symmetry of the thermal lens, which is cylindrically symmetric and approximately parabolic for the macroscopic variants, while it is spherically symmetric for the TL. Within the ABCD framework, ray-transfer matrices which encode the action of optical elements are used to transform Gaussian beams. Central to the evaluation of the relative PT signal is thus the identification of the proper optical element representing the thermal lens. A thin lens along with its transfer matrix $M_f=\left\{1,0;-f^{-1},1\right\}$ fulfils this task satisfactorily. The results from the geometrical optics treatment of the thermal lens, i.e.\ the ray-solution to Fermat's principle given in eq.\ \eqref{eq:Rutherford_Photon}, are then used to find an expression for the focal length of the thermal lens. However, the reason why this should not be expected to deliver perfect results are the dramatic spherical aberrations which characterize the thermal lens: The focal length of a macroscopic TL depends on the distance $b$ of the ray to the optical axis, $f_{\infty}\left(b\right) \approx b^2 n_0/\left[2\Delta n R\right]$. It varies quadratically with $b$ and, consequently, diverges as $b$ grows large. However, as a focused laser beam of finite extent probes this lens, large values of $b$ are not realized. Assigning a reasonable estimate of an effective squared impact parameter $b_{\rm eff}^2=\omega^2\left(z_p\right)$, one finds an agreement around $z_p=0$ with the paraxial solution in the diffraction model \eqref{eqPhiApproxDiffrOnAxis}. For a Gaussian beam located at a distance $z_p$ relative to the position of the TL the effective focal length $f_{\rm eff}$ should then follow
\begin{align}
f_{\rm eff}\left(z_p\right)\approx \frac{n_0}{\Delta n} \frac{\omega^2_0}{2R}\left[\frac{z_p^2}{z_R^2}+1\right].\label{eq:feff}
\end{align}
The strength of the TL as experienced by the probing beam is inversely proportional to the focal length and thus directly proportional to the refractive index contrast $\Delta n/n_0$. For the common case of materials with a negative thermorefractive coefficient $\partial_T n<0$ the focal length is negative, signifying a divergent lens. Using a concatenation of ray-transfer matrices to model the situation in PT microscopy, and transforming the Gaussian beam parameter of the probe beam one finds the far field beam waists
\begin{equation}
\frac{1}{\omega^2\left(z\right)}=\frac{1}{\omega_0^2}\frac{z_R^2}{z^2},\quad \frac{1}{\omega^2_{\rm TL}\left(z\right)}=\frac{1}{\omega_0^2}\frac{z_R^2}{z^2}\left[1+\frac{2z_p}{f_{\rm eff}}\right],\label{eq:ABCDbeamwaists}
\end{equation}
for the lenses and non-lensed probe beam. The resulting signal may be inferred from these assuming that the on-axis signal scales with the inverse beam waists squared, i.e.\ taking $\Phi=\left[\omega^{-2}_{\rm TL}-\omega^{-2}\right]/\omega^{-2}$, yielding
\begin{align}
\Phi\left(z_p\right)=2z_p/f_{\rm eff}.\label{eq:Phif}
\end{align}
On intuitive grounds, this is probably the most cogent argument why the contrast $\Delta n/n_0$ must linearly enter the PT signal at a fixed position of the probing beam waist. This conclusion is not corrupted by the neglect of the spherical aberrations or the full time-dependence which add to the details of the (thermal) lens as the corresponding calculations show. Considering a time-dependent focal length, the above expression may be generalized to provide the signal's frequency behaviour via the demodulation of the focal length at some fixed distance as
\begin{equation}
\Phi_{\rm sin}\propto \frac{1}{T_\Omega}\int_{0}^{T_\Omega}\sin\left(\Omega t\right) \frac{1}{f\left(t\right)}\mathrm{d}t\label{eq:nrSin},
\end{equation}
with a similar expression for the in-phase component, i.e.\ analogously to eq.\ \eqref{eq:PhiCos} and \eqref{eq:PhiSin}. Returning for now to the resulting PT signal for low modulation frequencies, one obtains:
\begin{equation}
\Phi\left(z_p\right) = \frac{2 P_{0,h} \sigma_{\rm abs} \left[\partial_T n\right]}{\lambda \pi \kappa\,\omega^2_{0,h}} \left[\frac{I_h\left(z_p\right)}{I_{h,0}}\right] \left[\frac{z_p}{z_{R}}\right] \left[\frac{I_d\left(z_p\right)}{I_{d,0}}\right].\label{eqn:PTSignal}
\end{equation}
The signal is seen to equal the product of the two beam's intensity profiles along the axial direction, which are Lorentzian in shape and possibly offset by $\Delta z_f$, and the scaled axial coordinate $z_p/z_R$. Despite the appearance of the Lorentzian probe beam profile times the axial coordinate instead of the inverse tangent found in the paraxially correct expression, Eq.\ \eqref{eqPhiApproxDiffrOnAxisd0}, the current expression Eq.\ \eqref{eqn:PTSignal} satisfactorily describes the signal magnitude and trends, see Fig.\ \ref{FigGaussABCD}. Indeed, both expressions coincide up to second order in the beam displacement. Also note that the maximum signal again scales with $\Delta n$ in the final expression \eqref{eqn:PTSignal} at some offset $z_p\sim z_R$.

\paragraph{frequency dependence for $R_{\rm th}/\omega_0\gg 1$}
The full time-dependent thermal lens at any frequency $\Omega$ and moment in time $t$ can be used to numerically find the focal length $f\left(t\right)$, which in turn may be used to find the in- and out-of-phase PT signals via Eq.\ \eqref{eq:nrSin}. In polar coordinates, the azimuthal component of Fermat's equation \eqref{eq:Fermat} shows the conservation of what can be referred to as the optical angular momentum. For each ray incident with an impact parameter $b$ it reads $L_z=r^2\phi'=n_0 b$. An integration the second equation in \eqref{eq:Fermat} can then be done, using $r'=\mathrm{d}r/\mathrm{d}s=\phi' \mathrm{d}r/\mathrm{d}\phi$:
\begin{equation}
\int_{\pi}^{\phi_0}\mathrm{d}\phi=\phi_0\left(t\right)-\pi=\int_{\infty}^{r_m}\frac{L_zr^{-2}\mathrm{d}r}{\sqrt{n^2\left(r,t\right)-L_z^2r^{-2}}}\label{eq:thetat},
\end{equation}
where $\phi_0$ and $r_m$ are the angle and radius of closest approach, respectively. The latter one can be found by numerically solving $r_m^2\phi'^2=n^2\left(r_m,t\right)$, since $r'=0$ at $r=r_m$. A numerical integration then provides the resulting deflection angle $\theta_d\left(t\right)=2\phi_0\left(t\right)-\pi$, replacing Eq.\ \eqref{eq:costhetahalfRF} for the ideal TL. It defines the ray's outgoing asymptote and thereby determines the effective focal length for each impact parameter, 
\begin{equation}
f\left(t\right)=-b/\sin\left(\theta_d\left(t\right)\right)\label{Eq:fnrt}
\end{equation}

\begin{figure}[t]
\centerline{\includegraphics[width=1.0\columnwidth]{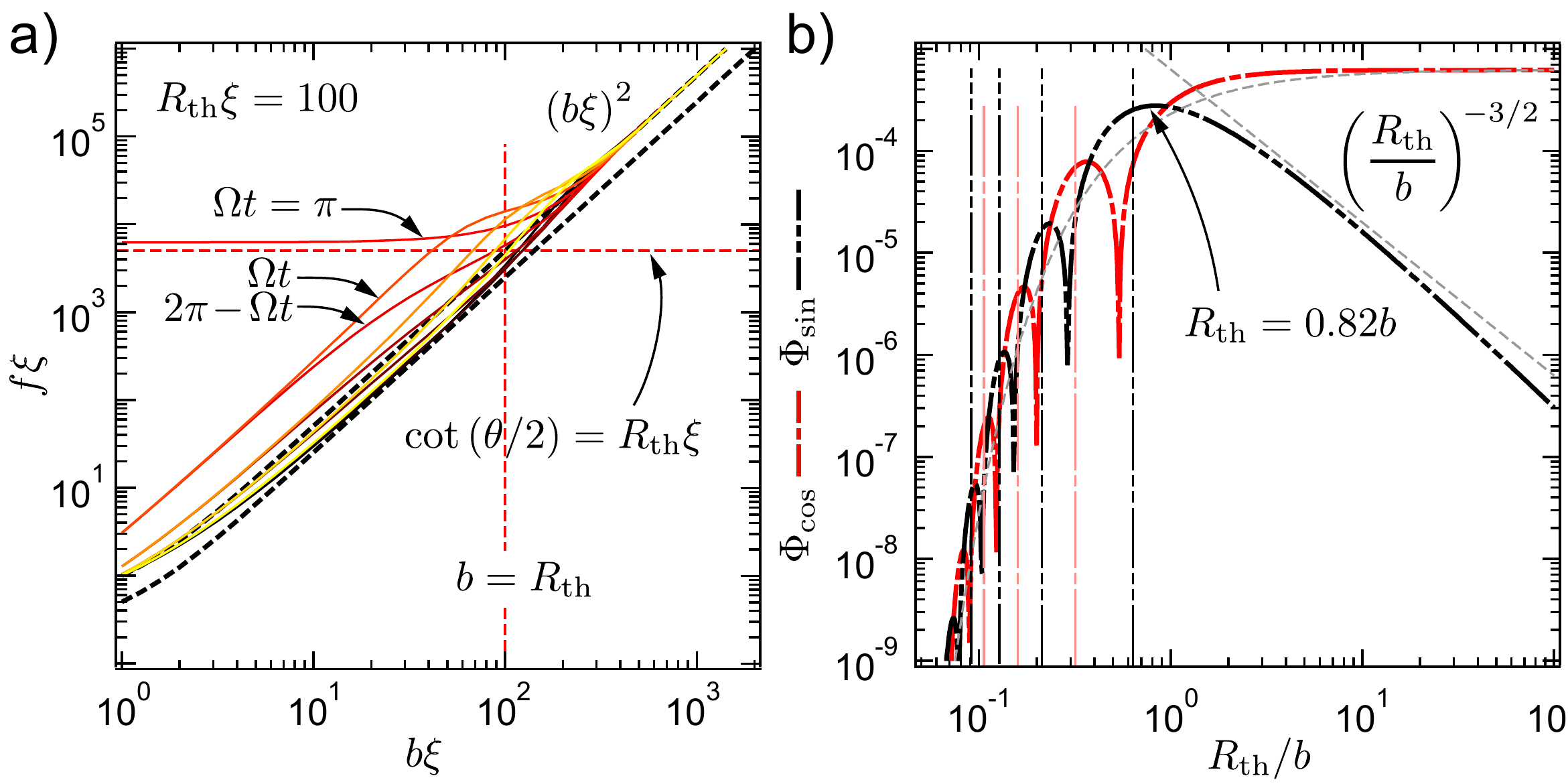}} %\textwidth
\caption{\textbf{a)} Focal length $\left[f\left(b\xi\right)\xi\right]^{-1}$, Eq.\ \eqref{Eq:fnrt}, for various instants $\Omega t\in\left[0,2\pi\right]$ (black to yellow). The dash black lines show $b^2\xi^2/2$ and $b^2\xi^2/4$. The plots were evaluated with $\xi=1$. \textbf{b)} In-phase and out-of-phase signals in the GOA limit via Eq.\ \eqref{eq:nrSin} in the weak lens limit ($b\xi=100$). The red dashed line shows the approximation $\pi \left(2/b^2\xi^2\right) \left[R_{\rm th}/b\,\right]^{-3/2}$. The grey dashed curve shows $\pi \left(2/b^2\xi^2\right)\exp\left(-b/R_{\rm th}\right)$. \label{FigAppendixGOA}} %$\nu=0.084$ wave-number $\bar{k}=14.45\mu{\rm m}^{-1}$
\end{figure}

The results of an evaluation of Eq.\ \eqref{eq:nrSin} and Eq.\ \eqref{eq:thetat} for $n\left(r,t\right)$ as in Eq.\ \eqref{eq:nr} are shown in Fig.\ \ref{FigAppendixGOA}. For impact parameters larger than the decay length, $R_{\rm th}/b\ll 1$, the time-dependent thermal lens is then equivalent to the ideal TL $n\left(r\right)=n_0+\Delta n R/r$. Indeed, the focal length closely follows the photonic Rutherford scattering result ($f\xi=-1/2\left[1+b^2\xi^2\right]$) in this regime, see the upper black dashed line in Fig.\ \ref{FigAppendixGOA}a). For impact parameters smaller than the decay length, $R_{\rm th}/b\gg 1$, the focal length approximately varies over time between nearly zero and the value corresponding to an ideal lens of twice the strength, see the lower black dashed line. Curves corresponding phases $\Omega t$ and $2\pi-\Omega t$ merge to a single one for very small values of $b\xi \ll R_{\rm th}\xi$. This reflects the fact that the term $\cos\left(\Omega t-r\xi/R_{\rm th}\xi\right)$ in the refractive index profile Eq.\ \eqref{eq:nr} is then be equal in both cases, and therefore the experienced deflecting gradient close to the coordinate center will be equal as well. The corresponding rays effectively propagate through a similar refractive index profile which is close to a fully modulated ideal TL $n\left(r\right)=n_0+\left[1+\cos\left(\Omega t\right)\right]\Delta n R/r$. The curve for $\Omega t=\pi$ shows a constant minimal deflection. The corresponding rays probe a refractive index profile for which the perturbation is canceled close to the center.  The constancy of the focal length can be understood as a consequence of Snell's law for the rays impinging nearly perpendicular to the remaining perturbation at $r\approx R_{\rm th}$: Considering Snell's law for a ray with an angle $\Psi=\arcsin\left(b/R_{\rm th}\right)$ to the normal on the sphere of radius $R_{\rm th}$, where it may be though to experience a refraction by a discontinuity, one may write $n_0 \sin\left(\Psi\right)=\left[n_0+\Delta n'\right]\sin\left(\Psi'\right)$ with $\Delta n'=- n_0/\xi R_{\rm th}$ being the perturbation at $r=R_{\rm th}$. The deflection angle is then $\theta=2\left[\Psi'-\Psi\right]$, which for small $b/R_{\rm th}$ amounts to $\theta \approx 2b/\xi R_{\rm th}$. This in turn leads to the constant value of $f\xi\approx R_{\rm th}^2 \xi^2/2$, see the horizontal red dashed line in Fig.\ \ref{FigAppendixGOA}a). This value coincides with the focal length experienced on the ideal TL for $b=R_{\rm th}$, i.e.\ $\cot\left(\theta/2\right)=R_{\rm th}\xi$. The exact value is obtained for $b\approx 1.126 R_{\rm th}$. The curve corresponding to the phase $\Omega t=0$ (black solid) show the strongest lens and thus a minimal focal length, corresponding to a lens which is nearly completely modulated and at its maximum contrast.

Overall, the spherical aberrations of a weak thermal lens ($b\xi\gg 1$) deviate markedly from those of the ideal TL $f\propto b^2$ only for $R_{\rm th}/b\approx 1$. It is also in this regime where a significant out-of-phase behaviour in the inverse focal length and thereby in the PT signal, Eq.\ \eqref{eq:nrSin}, appears. In fact, a maximum out-of-phase signal is found for $R_{\rm th}\approx 0.82 b$, see Fig.\ \ref{FigAppendixGOA}b). The oscillations for larger impact parameters where $R_{\rm th}/b< 1$ occur at the extrema of the corresponding refractive index field wave $\sin\left(-r/R_{\rm th}\right)$, i.e.\ at $b/R_{\rm th}=\left[2n+1\right]\pi/2$ with $n=0,1,2,\dots $, see the grey dashed-solid vertical lines in the figure. Similarly, the in-phase component shows local maxima for impact parameters $b/R_{\rm th}=n\pi$ such that $\cos\left(-r/R_{\rm th}\right)$ is maximal, see the red double-dashed solid lines. The amplitude of both components follows an exponential decay, $\pi\left[2/b^2\xi^2\right]\exp\left(-b/R_{\rm th}\right)$, see the bent grey dashed line. Here, the factor $\pi$ accounts for the average of $\cos^2\left(\Omega t\right)$. These oscillations could likely also be exploited to quantify the thermal diffusivity by an offset dependent measurement of the single particle Rutherford scattering microscopy signal using a quadrant or balanced photodiode. Note that this regime does \textit{not} correspond to the high-frequency PT lensing signal $\left[R_{\rm th}/\omega_0\right]^2$ which is a phenomenon of scattering in wave-optics outside the realm of geometrical optics.

Returning to the case of $R_{\rm th}/b>1$, which is relevant for the conventional single particle PT signal at low frequencies, the power-law decay for of the out-of-phase signal, here defined via Eq.\ \eqref{eq:nrSin}, $\Phi_{\rm sin}\approx \pi \left[2/b^2\xi\right] \left[R_{\rm th}/\omega_0\right]^{-3/2}$ is thus seen as a remaining lensing phenomenon in the geometrical optics regime. The peak in the out-of-phase signal at $R_{\rm th}\approx \omega_0$, found before in the rigorous GLMT framework, may thus be taken to correspond to the peak in Fig.\ \ref{FigAppendixGOA}b).

Again one may consider an effective impact parameter close to the local beam-waist $b\sim \omega_0$ to find a semi-quantitative approximation of both components of the photothermal signal at low frequencies. In fact, a good agreement was found \cite{SelmkeApplPhysLett2014} using $b=\omega\left(z_{R}\right)=\sqrt{2}\omega_{0}$, i.e.\
\begin{equation}
\Phi_{\rm sin}\left(z_{R,d}\right)=\frac{P_A\sigma_{\rm abs}\partial_T n}{1+z_{R}^2/z_{R,h}^2}\frac{1}{\kappa\omega_{0,h}^2\lambda}\left[\frac{R_{\rm th}}{\sqrt{2}\omega_0}\right]^{-3/2}\label{eq:SinApprox}
%\frac{\Delta n}{1+z_{R,d}^2/z_{R,h}^2}\frac{2\pi^2 R}{\lambda_d}\left[\frac{R_{\rm th}}{\sqrt{2}\omega_0}\right]^{-3/2}\label{eq:SinApprox}
\end{equation}

%The effect of modulated heating may be anticipated by considering a finite thermal lens which is sharply cut off at a distance $R_{\rm th}$. Such a thermal lens may be seen as a crude approximation the the expression $n\left(r,t\right)$, eq.\ \eqref{eq:nr}. Then, the focal length increases according to $f_{R_{\rm th}}/f_\infty \approx 1+b^2/2R_{\rm th}^2$. This may be seen to indicate the onset of the signal decay $\Phi\propto R_{\rm th}^2$ which is found using the rigorous GLMT model.

\paragraph{finite aperture dependence}
From Eq.\ \eqref{eq:ABCDbeamwaists}, the angular structure of the PT signal, $\Phi\left(\theta\right)$, is here predicted to be the difference between the original Gaussian beam and the modified Gaussian beam which is either collimated or diverged. These predictions agree qualitatively and semi-quantitatively with the observed trends. 

In the current description the aperture dependence of the signal may still be estimated using the notion of a broadened probe beam. To this end, the far field Gaussian field distributions described by a beam waist, either transformed or not, should be integrated over the corresponding angular domain of detection. Using the results for the beam-waists and computing in the zeroth order in the small quantity $1/f_{\rm eff}$, a focal-length and position $z_p$ independent probe beam characteristic function $F$ is obtained which scales the on-axis signal.
\begin{equation}
F=\frac{\vartheta_r\left(\theta_{\rm max}\right)e^{\vartheta_r\left(\theta_{\rm min}\right)} - \vartheta_r\left(\theta_{\rm min}\right)e^{\vartheta_r\left(\theta_{\rm max}\right)}}{e^{\vartheta_r\left(\theta_{\rm max}\right)}-e^{\vartheta_r\left(\theta_{\rm min}\right)}}\label{eq:ABCDFfinal}.
\end{equation}
However, the diffraction model (E) captures the details of the signal decay with increasing detection aperture somewhat better, see Fig.\ \ref{FigGaussABCD}.

\section{PT microscopy involving phase transitions}
A further model-system has been found in liquid crystal. These materials undergo phase-transitions and therefore provide a large change in their refractive index upon heating. Therefore, they may be used to enhance the photothermal signal \cite{Lounis2012LC,Link2012LC}. However, the pumping frequency $\Omega$ has been shown to be a crucial factor since the latent heat must be provided and transported away at the generated phase boundary, which is a process that occurs non-instantaneously \cite{Heber2013}. Also, the optical scattering problem is no longer a simple thermal wave but also entails a discrete moving boundary and therefore a moving Mie-scatterer or ball-lens. The so-called Zharov splitting observed in bulk absorption PT microscopy is related to the complex behaviour of the liquid crystal's driven phase transition \cite{Zharov2012,MertiriApplPhys2012,MertiriACSNano2014}.

\section{Conclusion}
In recent years photothermal microscopy has become a valuable tool in condensed-matter physics and plasmonics. By employing an approach which may be regarded as heterodyne, its detection sensitivity reaches down to single absorbing molecules. Modelling of the transmission signal in single particle photothermal microscopy allows to quantitatively assess absorption cross-sections of individual particles and thermal diffusivities of the surrounding medium. Mastering of the PT technique will establish it as a standard tool for the imaging and characterisations of particles and host-systems alongside the well-established fluorescence microscopy techniques. %Further progress will likely give access to new thermal-wave related phenomena or phase-transition dynamics on a nanoscale. 

\end{document}